\documentclass[journal,12pt,onecolumn]{IEEEtran}

\usepackage{amssymb}
\usepackage{epsfig}
\usepackage{algorithmic}

\usepackage{fullpage}
\usepackage{cite}
\usepackage[cmex10]{amsmath}
\usepackage{mdwmath}
\usepackage{mdwtab}
\usepackage{caption}
\usepackage{verbatim}
\usepackage[tight,footnotesize]{subfigure}

\usepackage[ruled,linesnumbered]{algorithm2e}
\renewcommand{\algorithmcfname}{ALGORITHM}
\SetAlFnt{\small}
\SetAlCapFnt{\small}
\SetAlCapNameFnt{\small}
\SetAlCapHSkip{0pt}
\IncMargin{-\parindent}

\newtheorem{theorem}{Theorem}[section]

\newenvironment{proof}[1][Proof]{\begin{trivlist}
\item[\hskip \labelsep {\bfseries #1}]}{\end{trivlist}}

\newcommand{\qed}{\nobreak \ifvmode \relax \else
      \ifdim\lastskip<1.5em \hskip-\lastskip
      \hskip1.5em plus0em minus0.5em \fi \nobreak
      \vrule height0.75em width0.5em depth0.25em\fi}

\begin{document}

\title{Cooperative Topology Control with Adaptation for Improved Lifetime\\in Wireless Sensor Networks}
\author{\IEEEauthorblockN{Xiaoyu Chu and Harish Sethu}\\
\IEEEauthorblockA{Department of Electrical and Computer Engineering\\
Drexel University\\
Philadelphia, PA 19104-2875\\
Email: \{xiaoyu.chu, sethu\}@drexel.edu}
}

\maketitle
\thispagestyle{empty}
~\vskip 0.5in
\begin{abstract}
Topology control algorithms allow each node in a wireless multi-hop
network to adjust the power at which it makes its transmissions and
choose the set of neighbors with which it communicates directly, while
preserving global goals such as connectivity or coverage. This allows
each node to conserve energy and contribute to increasing the lifetime
of the network. In this paper, in contrast to most previous work, we
consider (i) both the energy costs of communication as well as the
amount of available energy at each node, (ii) the realistic situation
of varying rates of energy consumption at different nodes, and (iii)
the fact that co-operation between nodes, where some nodes make
a sacrifice by increasing energy consumption to help other nodes reduce
their consumption, can be used to extend network lifetime. This paper
introduces a new distributed topology control algorithm, called the
{\em Cooperative Topology Control with Adaptation (CTCA)}, based on a
game-theoretic approach that maps the problem of maximizing
the network's lifetime into an ordinal potential game. We prove the
existence of a Nash equilibrium for the game. Our simulation results
indicate that the CTCA algorithm extends the life of a network by more
than 50\% compared to the best previously-known algorithm. We also
study the performance of the distributed CTCA algorithm in comparison
to an optimal centralized algorithm as a function of the communication
ranges of nodes and node density.
\end{abstract}

\newpage
\section{Introduction}\label{sec:introduction}
In wireless ad hoc networks, especially ad hoc sensor networks, the
battery life of each node plays a critical role in determining the
functional lifetime of the entire network. When a node exhausts its
limited energy supply, it may fail to reach nearby nodes leading to a
disconnected network and disabling some essential
communications. Without energy, the node will also fail to continue
the environmental monitoring activities essential to the functional
operation of the system. Adding redundant nodes in the network may
extend the functional lifetime but it is ultimately a less
cost-effective approach. In this paper, we consider the problem of 
extending the lifetime of a network using a new adaptive
game-theoretic approach.

Topology control is among the better-known approaches to
conserving energy and prolonging a network's functional life. In a
topology control algorithm, each node adjusts the power at which it
makes its transmissions to reduce the energy consumption to only what
is needed to ensure topological goals such as connectivity or
coverage. Examples of topology control
algorithms include Directed Relative Neighborhood Graph (DRNG) \cite{LiHou2005-1313},
Directed Local Spanning Subgraph (DLSS) \cite{LiHou2005-1313}, Step Topology Control (STC) \cite{SetGer2010} and Routing Assisted Topology Control (RATC) \cite{KomMac2009}. In most traditional algorithms, the topology of the network is
determined at the very beginning of the life of the network where the
only consideration for each node is to reduce its transmission power
while keeping the graph connected. After the execution of one of these
algorithms, each node will transmit at the selected power level until
it eventually runs out of energy. However, depending on the location
of a node in relation to others, some nodes may end up with a much
larger communication radius, and therefore a much larger transmission
power, than some others. This uneven distribution of the assigned
transmission powers may result in an unbalanced energy consumption at
the nodes, leading to some nodes exhausting their energy far sooner
than some others. Such a scenario can end the functional life of the
network earlier than necessary. This highlights two weaknesses of
these algorithms: they are not adaptive to different rates of energy
consumption on different nodes and they do not allow
cooperation between nodes to extend the network lifetime. Each of
these weaknesses is addressed by the algorithm proposed in this paper:
{\em Cooperative Topology Control with Adaptation (CTCA)}.

\begin{figure}[!h]
\begin{center}
    \subfigure[{The initial topology of the network where no node can reduce its transmission power.}]{
        \includegraphics[width=1.5in]{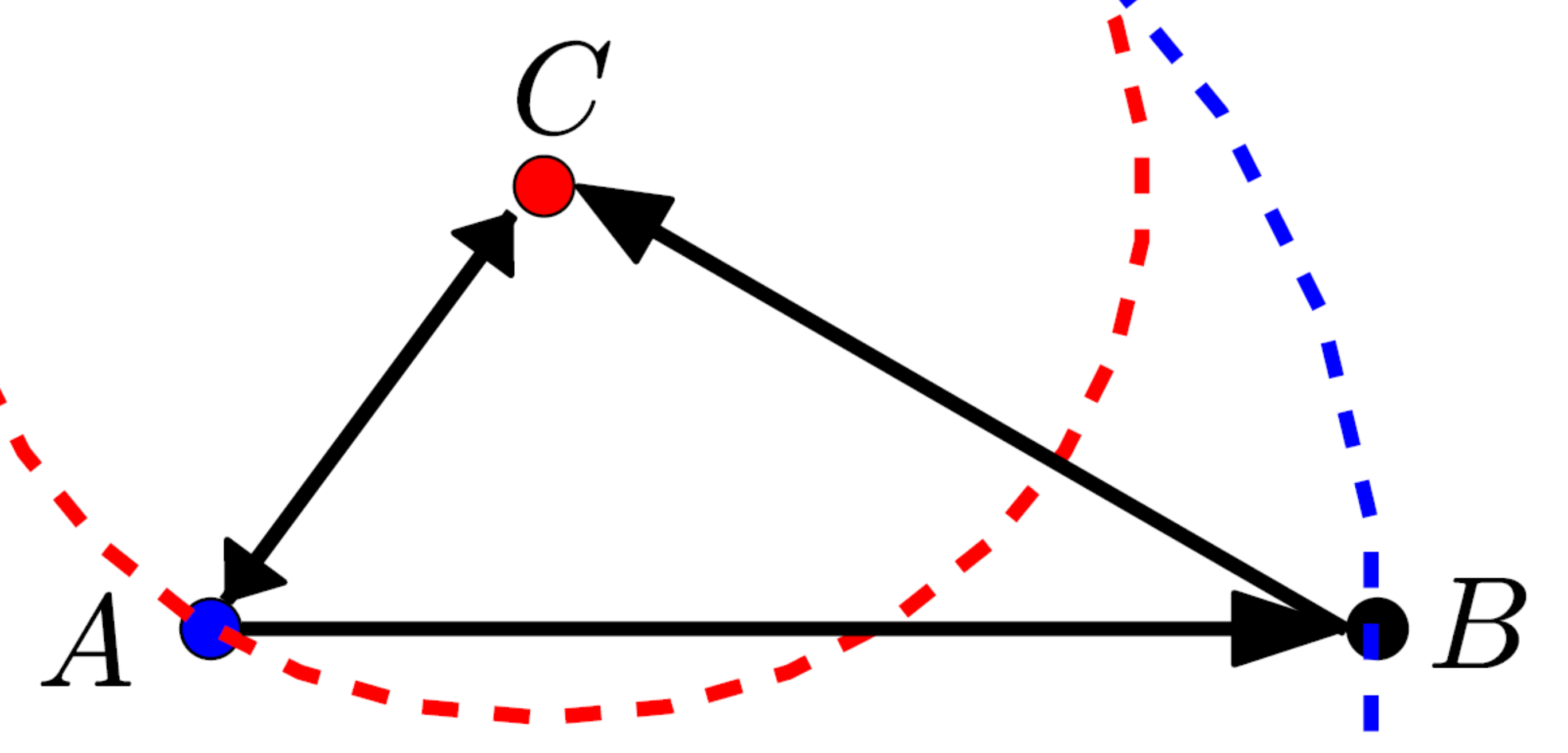}
        \label{fig:1-before}
        }\hskip0.2in
    \subfigure[{Node $C$ chooses to increase its transmission power so
      as to directly connect to $B$.}]{
        \includegraphics[width=1.5in]{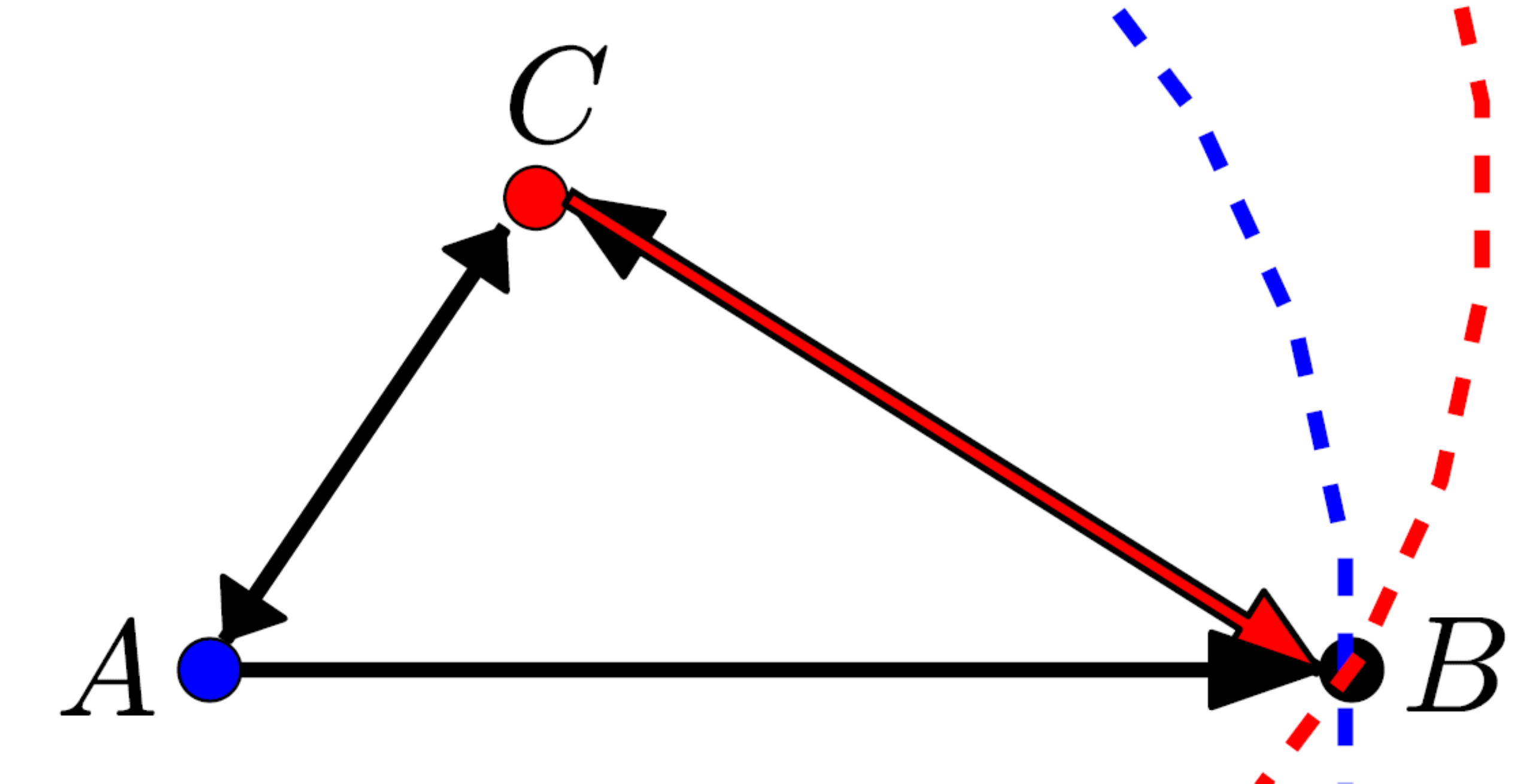}
        \label{fig:1-after}
        }\hskip0.2in
    \subfigure[{Node $A$ can now reduce its transmission power to
      directly connect only to $C$.}]{
        \includegraphics[width=1.5in]{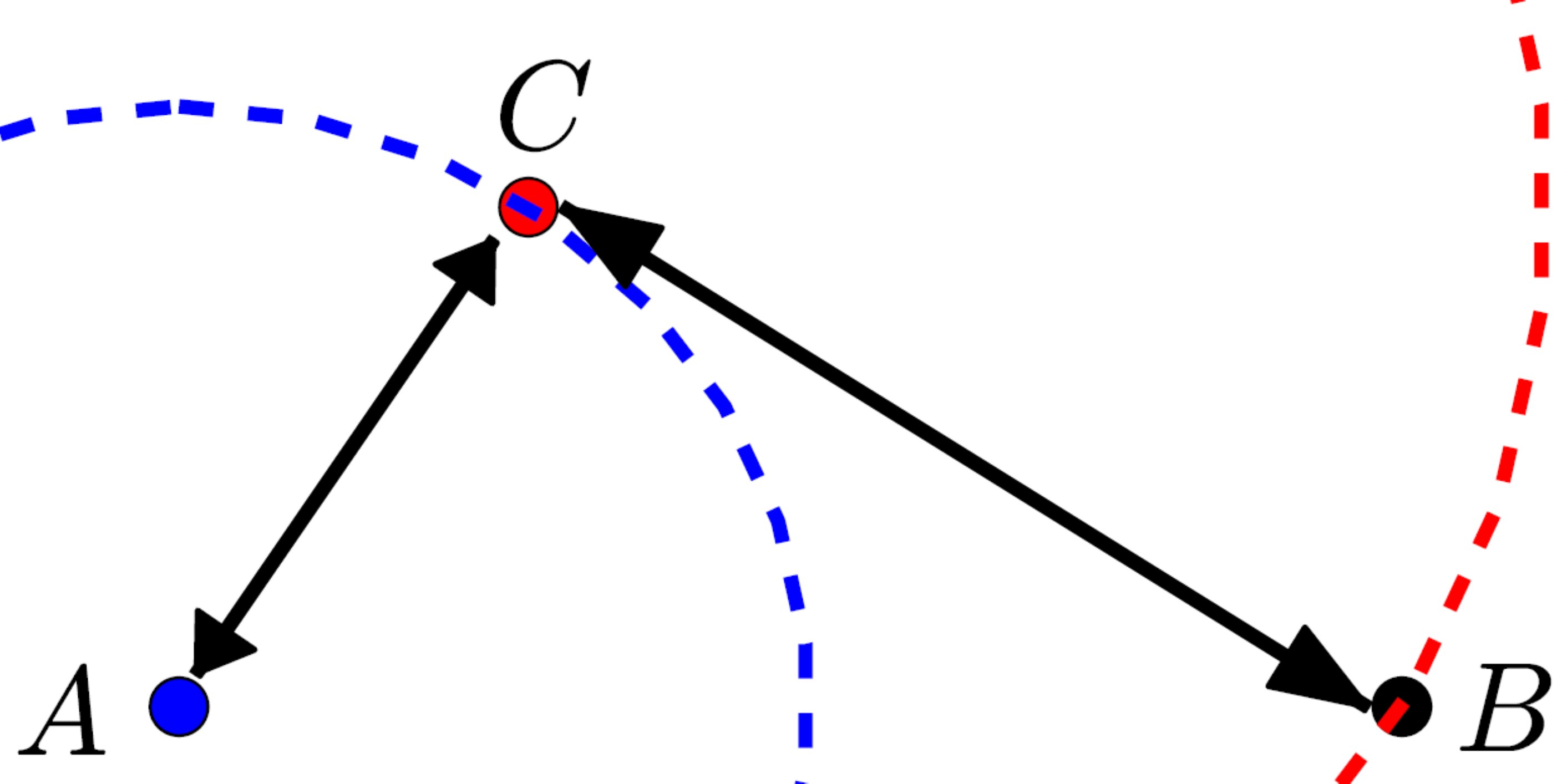}
        \label{fig:1-after-1}
        }
    \caption{An example illustrating cooperative topology control.}
    \label{fig:1-improve}
\end{center}
\end{figure}
We illustrate the principle of cooperative topology control with a
simple toy example shown in Fig. \ref{fig:1-improve}. Suppose
Fig. \ref{fig:1-before} illustrates the result of a topology
control algorithm, where no node can reduce its transmission power
unilaterally without disconnecting the graph. In this figure, the
presence of an edge from one node, say $A$, to another node, say $B$,
implies that $A$ can transmit at a power level sufficient to reach
node $B$. The communication radius of each node is shown by the dashed
arcs. Assuming all nodes start out with the same energy supply and
make transmissions at the same rate, we note that node $A$ has the
largest energy cost and thus has the shortest lifetime. Node $C$, on
the other hand, has the smallest transmission power, and therefore,
has the longest lifetime. Traditional topology control algorithms
discussed above will lead to the situation in Fig. \ref{fig:1-before}
ending the functional life of the network when node $A$'s energy is
exhausted even though node $C$ would have plenty of remaining
energy. Fig. \ref{fig:1-after} illustrates a topology where node
$C$ increases its transmission power so that it can now reach node $B$
directly. Now, node $A$ is able to reduce its transmission power to
only be directly connected to node $C$, as shown in
Fig. \ref{fig:1-after-1}. This involves node $C$ making a sacrifice
by increasing the power at which it makes its transmissions in order
to allow node $A$ to reduce its transmission power, thus extending
the life of node $A$ and of the network.

The actual power consumption for sending and receiving a data
packet varies significantly depending on the radio environment of the
space where the sensor nodes are located and also the electronics of
the devices. The log-distance path
loss model based on the path loss as a logarithmic function of the
distance $d$ has been confirmed both theoretically and by measurements
in a large variety of environments \cite{Rap2002,GenVai2007,SunAky2010}. In this
model, the path loss at distance $d$, $PL(d)$ is expressed as: 
\[
PL(d) = PL(d_0) + 10 \gamma \log_{10}( d/d_0 )
\]
where the constant $d_0$ is an arbitrary reference distance and
$\gamma$ is called the path loss exponent. This implies that the
energy consumed to make a transmission across a distance $d$ is
proportional to $d^\gamma$. Since $\gamma$ ranges from 2.5 to
6 in most real environments \cite{Rap2002}, especially over longer
distances, a single transmission over distance $d$ often consumes more
energy than two transmissions each over distance $d/2$. This motivates
the goal of most topology control algorithms to choose multiple
smaller hops in place of a single longer hop with the intent to reduce
overall energy consumption. While device electronics can sometimes be
such that choosing smaller hops---especially at smaller
distances---does not always guarantee lower energy consumption, there
is another good reason to choose smaller hops: reducing interference
in all communications. Therefore, a general goal of a topology control
algorithm is to achieve lower transmission powers for all the sensor
nodes in order to reduce both energy consumption and interference
\cite{ChiHan2012}. In other words, the topology illustrated in
Fig.\,\ref{fig:1-after-1} is more desirable.  

In this paper, we employ game theory to facilitate such topology
control that allows cooperation between nodes as illustrated in
Fig. \ref{fig:1-improve}. Our approach is through developing an
ordinal potential game \cite{MarArs2009,KomMac2009,MonSha1996} into which our
problem can be mapped, so that all nodes pursue a localized strategy
that can be expressed through a single global function, or the global potential function. Our
approach also allows an adaptive strategy so that a node does not end up
with the same power level through its entire lifetime. This is
significant to extending the network lifetime because it is almost
always the case that different nodes consume energy at different
rates. Our approach to allowing adaptation is through incorporating
the energy remaining on the nodes in the neighborhood into the
decisions made by each node. Since this remaining energy changes over
the life of a network, our topology control algorithm adaptively
adjusts the power levels at each node. This constantly keeps shifting
energy consumption from nodes with less energy reserves to those with
more energy reserves, thus extending the life of the network.

\subsection{Problem Statement}\label{sec:problemStatement}

Given a wireless sensor network, let graph $G(t)=(N,E(t))$ represent its
topology at time $t$, where $N_i\in N$ represents a node within
the network, and $(N_i,N_j)\in E(t)$ represents the fact that node $N_j$
is within node $N_i$'s communication radius and can hear from $N_i$
directly at time $t$.
Assume $G(t)$ is a connected graph at time $t$.
Topology control algorithms have traditionally
emphasized preserving connectivity as a constraint while pursuing the
goal of reduced energy consumption at each node.


However, depending on the type of application for which an ad hoc
sensor network is deployed, it is possible that a network is
functional even if a certain subset of nodes runs out of energy
\cite{DieDre2009,DenHan2005,DuaLiu2002}. The functional lifetime of a
network, therefore, depends on the application in use and
consequently, there is some debate on how best to define the
functional lifetime of a network. In an ad hoc
sensor network with a non-hierarchical topological organization, one
may assume an application-dependent parameter, $c$, to define the
functional lifetime as the length of time the network topology
possesses at least one connected component with $n-c$ or more nodes
(where $n$ is the total number of nodes in the network). When $c=0$,
the functional lifetime of the network is the length of time $G(t)$ is
a connected graph, which is until any node is disconnected or runs out
of energy.

We find that a definition of functional lifetime using $c=0$ is a more
versatile one for two reasons: firstly, on any application, there may
be some crucial nodes which, when they die, can disable the
functionality of the network; secondly, a definition based on the
$c=0$ case can form the foundation of greedy algorithms designed to
extend functional lifetime for $c>0$.

In this paper, therefore, 
we consider the functional life of the network to have
ended when one of the following two cases occurs:
\begin{itemize}
\item Case 1: A node reduces its current transmission power in order
  to save energy, but becomes unable to reach certain nodes and,
  consequently, loses connection from part of the network.
\item Case 2: A node runs out of energy, thus getting disconnected
  from the rest of the network.
\end{itemize}

If Case 1 happens, the communication links whose removal caused
the network to become disconnected can be restored back into the
network to restore the functional life of the network. On the other
hand, if Case 2 happens, the network's functional lifetime cannot be
extended in any way. Therefore, to improve the lifetime of the
network, (i) Case 1 should be avoided by always ensuring connectivity in
the assignment of power levels to the nodes, and (ii) Case 2 should be
pushed as far into the future as possible by reducing the rate of
energy consumption at the node that is estimated to have the smallest
remaining lifetime. The problem can now be defined as one of
periodically reassigning the power at which each node makes its
transmissions so that the first occurrence of either Case 1 or Case 2
is pushed as far ahead in time as possible.

\subsection{Contributions and Organization}

Section \ref{sec:related_work} reviews the related work on approaches
that have been employed to increase a wireless sensor network's
lifetime through topology control. Section \ref{sec:analysis} analyzes the rationale behind
the approach used in this paper and presents a few definitions and lays out
the foundational concepts for the game-theoretic approach used in this
paper. Section \ref{sec:game} proves the existence of a Nash
equilibrium for the ordinal potential game used to map our problem. Our proof
is based on showing that the difference in individual payoffs for each
node from unilaterally changing its strategy and the difference in
values of the global potential function have the same sign.

The pseudo-code for the Cooperative Topology Control with Adaptation
(CTCA) algorithm is presented in Section \ref{sec:pseudo_code}. A
simulation-based evaluation of its performance and a comparative
analysis with other topology control algorithms are described in
Section \ref{sec:CTCA_performance}. Our results show that the CTCA
algorithm extends the life of a network by more than 50\% compared to
the best previously-known algorithm. This section also compares the
topology delivered by the distributed CTCA algorithm to the optimal
solution obtained using a centralized algorithm. We also study the
dependence of the performance of CTCA in relation to the optimal on
the communication ranges of nodes and on the node density.

Section \ref{sec:conclusion} concludes the paper.

\section{Related Work}\label{sec:related_work}

The task of extending the life of a wireless sensor network can be
tackled through multiple complementary ways involving routing
protocols, medium access strategies or any of several other
protocols that facilitate network operations. In this section, we will
discuss only the approaches most related to this paper; that is,
approaches based on changing the topology of the network by individual
nodes changing the power levels at which they make their transmissions
while preserving network connectivity.

Traditional topology control algorithms such as Small Minimum-Energy
Communication Network (SMECN)\cite{LiHal2001}, Minimum Spanning Tree
(MST) \cite{LiHou2002}, DRNG\cite{LiHou2005-1313},
DLSS\cite{LiHou2005-1313} and STC\cite{SetGer2010} usually start the
topology control process with each node transmitting at its maximum
transmission power to discover all of its neighbors. Local
neighborhood and power-level information is next exchanged between
neighbors. The minimum transmission power of each node such that the
graph is still connected is later computed at each node without
further communication between nodes. The Weighted Dynamic Topology
Control (WDTC) \cite{SunYua2011} algorithm improves upon the work of
MST, and considers the remaining energy of each node
in addition to the energy cost of communication across each pair of
nodes. The algorithm, however, forces bidirectional communication
between each pair of nodes and, in addition, requires periodic
communication by each node at its maximum possible power level.
Other related algorithms seek to offer a robust topology where the
graph can stand multiple channel failures; for example, a
$k$-connected graph is sought in \cite{HajImm2007,MiyNak2009} and a
two-tiered network in \cite{PanHou2003}.

Other topology control algorithms may require communication between
nodes throughout the topology control process. One typical example
is the work described in \cite{KomMac2006}, which is based on a
selfish game on network connectivity to help reduce the transmission
power on each node. By offering a utility function which indicates a
high profit if the node's transmission power is small and a low
profit if the node's transmission power is large, each node
selfishly reduces its transmission power to maximize its profit. On
the other hand, if the node has reduced its transmission power to
such an extent that the graph becomes disconnected, the profit of
each node becomes 0. This algorithm was later improved in
\cite{KomMac2009}, where the requirement of global information (to
establish connectivity) is eliminated and a distributed
topology control algorithm is proposed.
Among the first works on using game theory in topology control
problems is \cite{EidKum2003} which gives tight bounds on worst-case
Nash equilibria for a game in which the network is required to
preserve connectivity. However, this study only considers selfish
nodes which try to minimize their energy consumption without
considering potential sacrifices nodes can make (by expending more
energy) to extend a network's lifetime.

Another class of topology control algorithms is represented by the
work reported in \cite{RenMen2009}, where the authors provide a
decentralized static complete-information game for power scheduling,
considering both frame success rate and connectivity. Yet other
approaches to increasing the lifetime of a wireless sensor network
include grouping nodes into clusters to create a communication
hierarchy in which nodes in a cluster communicate only with their
cluster head and only cluster heads are allowed to
communicate with other cluster heads or the sink node
\cite{AbaAna2009,YouFah2004,KolPav2011,UstLin2011,VouAna2013,LiDel2013}. In the work of
\cite{XenKat2012}, the authors tried to assign sensor nodes with
different initial energy levels so that sensor nodes with high traffic
load will be assigned more energy than those with smaller loads. By
doing so, with the same amount of overall energy, the network's
lifetime may be extended.

If the network's lifetime is measured in terms of how many
transmissions can be made before the sensor nodes run out of energy,
then maximizing the network's lifetime can be interpreted as
maximizing the throughput of the network. In the work of
\cite{LuoIye2011}, the authors studied the relationship between
throughput of the network and its corresponding lifetime under an SINR
model. But they focus on a specific network setting where sensor
nodes' neighbors and the communication links are predetermined and the
topology of the network remains constant throughout the network's
lifetime.

A survey of topology control algorithms can be found in
\cite{MahMin2008,San2005} and a survey of the applications of game
theory in wireless sensor networks can be found in
\cite{MacTek2008,AziSek2013}.

The CTCA algorithm proposed in this paper is the first to use a
game-theoretic approach that also adapts to changes in the
remaining energy levels of nodes and which allows co-operative
behavior amongst nodes. As will be discussed in the
following sections, these features allow it to
extend the life of a network by more than 50\% compared to the
best previously-known algorithm.

\section{Definitions and Preliminaries}\label{sec:analysis}

In this section, we define terms and concepts that will enable us to
specify the localized goals that each node should pursue in order to
achieve the global goal of increased lifetime for the network.

Let $W_i(t)$ denote the amount of energy remaining at node $N_i$ at time
$t$. Let $p_i(t)$ denote the power at which node $N_i$ makes its
transmissions at time $t$. As an estimate of the additional length of
time before a node runs out of energy, we define the \emph{estimated
  lifetime} of node $N_i$ at time $t$, denoted by $L_i(p_i(t),t)$, as
the ratio between the amount of
remaining energy on the node at time $t$ and the power at which it
makes its transmissions at time $t$. That is, $L_i(p_i(t),t)=W_i(t)/p_i(t)$.
Note that the estimated lifetime may or may not accurately capture the
actual remaining lifetime of a node (because its transmission powers
may change later or its energy reserves may deplete slower/faster than
estimated.) When the context is clear, for brevity, we refer to the
estimated lifetime as just the {\em lifetime}.

In a system in which the rate of energy consumption is largely
balanced across the nodes (which is the goal of this paper as a means
to improve network lifetime), the node with the smallest estimated
lifetime is {\em likely} the one that determines the network's
lifetime. We consider the {\em estimated lifetime} of a network as the
estimated lifetime of the node with the smallest estimated
lifetime. If $N_i$ is the node with the smallest estimated lifetime
within the network, then it may be possible to improve the network's
lifetime by improving $N_i$'s estimated
lifetime. Fig. \ref{fig:1-improve} shows an example where node $A$ is
able to reduce its transmission power with help from node $C$, thus
increasing its estimated lifetime and likely the lifetime of the network.

\begin{figure}[!t]
\begin{center}
    \subfigure[{Node $N_3$'s initial state, where node $N_1$ is not capable of reducing its transmission power without disconnecting the graph.}]{
        \includegraphics[width=2.2in]{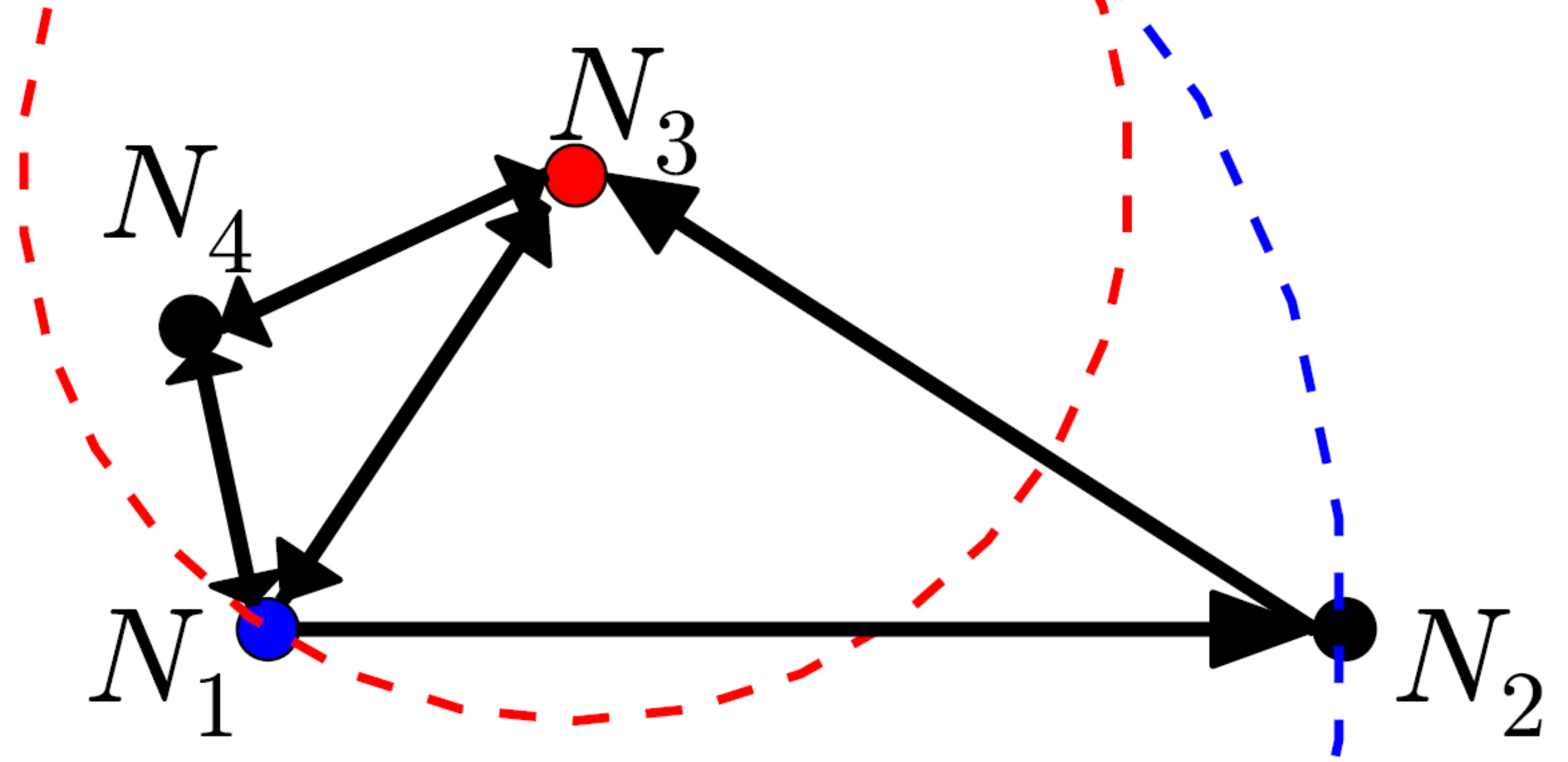}
        \label{fig:2-before}
        }\hskip0.5in
    \subfigure[{Situation 1: Node $N_3$ reduces its transmission power to its potential transmission power, but $N_1$'s lifetime cannot be improved.}]{
        \includegraphics[width=2.2in]{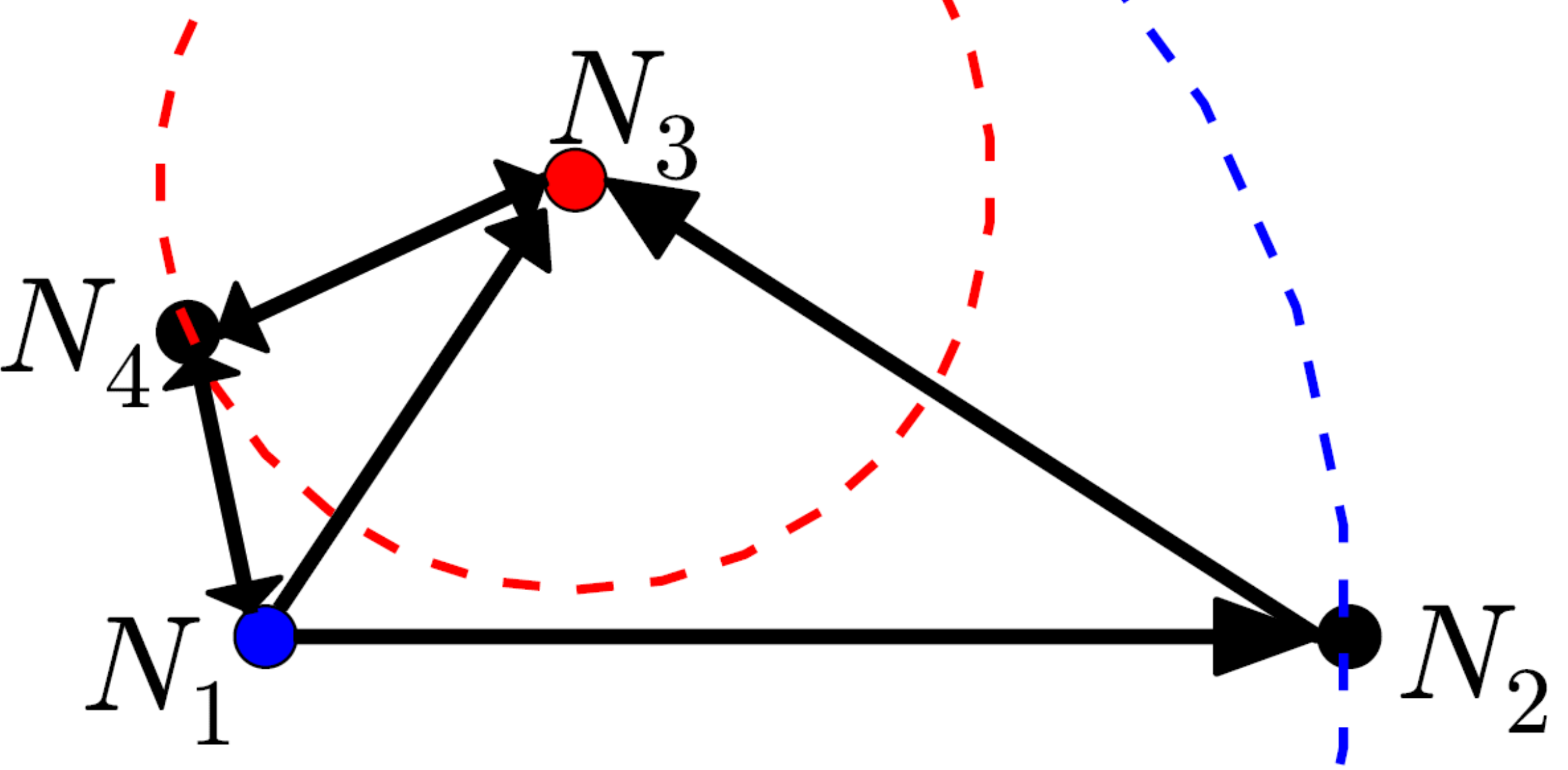}
        \label{fig:2-after-2}
       }\\
    \subfigure[{Situation 2: Node $N_3$ increases its transmission
      power to $p(N_3,N_2)$. Now, node $N_1$ is able to reduce its power without disconnecting the graph.}]{
        \includegraphics[width=2.2in]{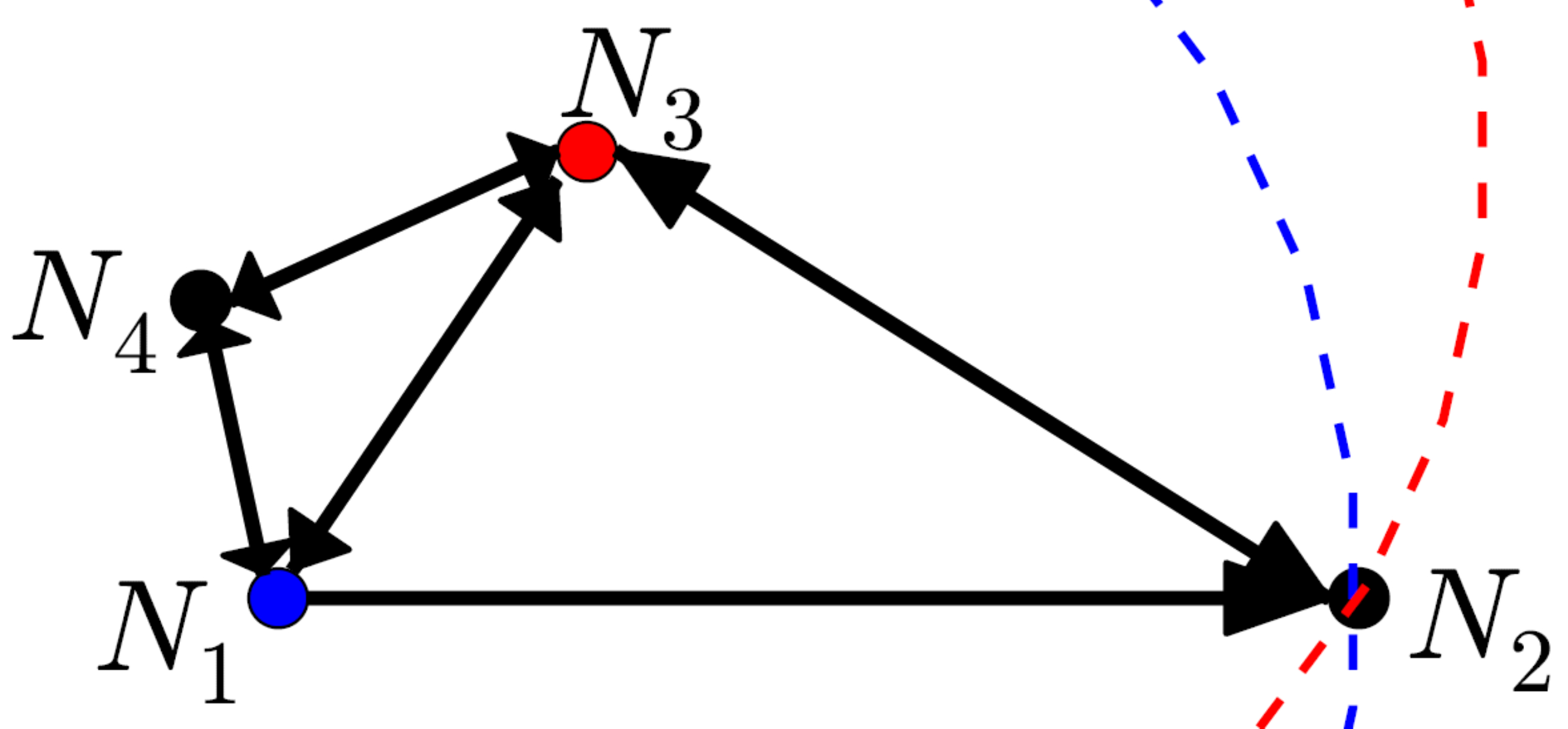}
        \label{fig:2-after-1}
        }\hskip0.5in
    \subfigure[{Node $N_1$ updates its current transmission power to
      its potential transmission power, and extends its lifetime and
      of the network. }]{
        \includegraphics[width=2.2in]{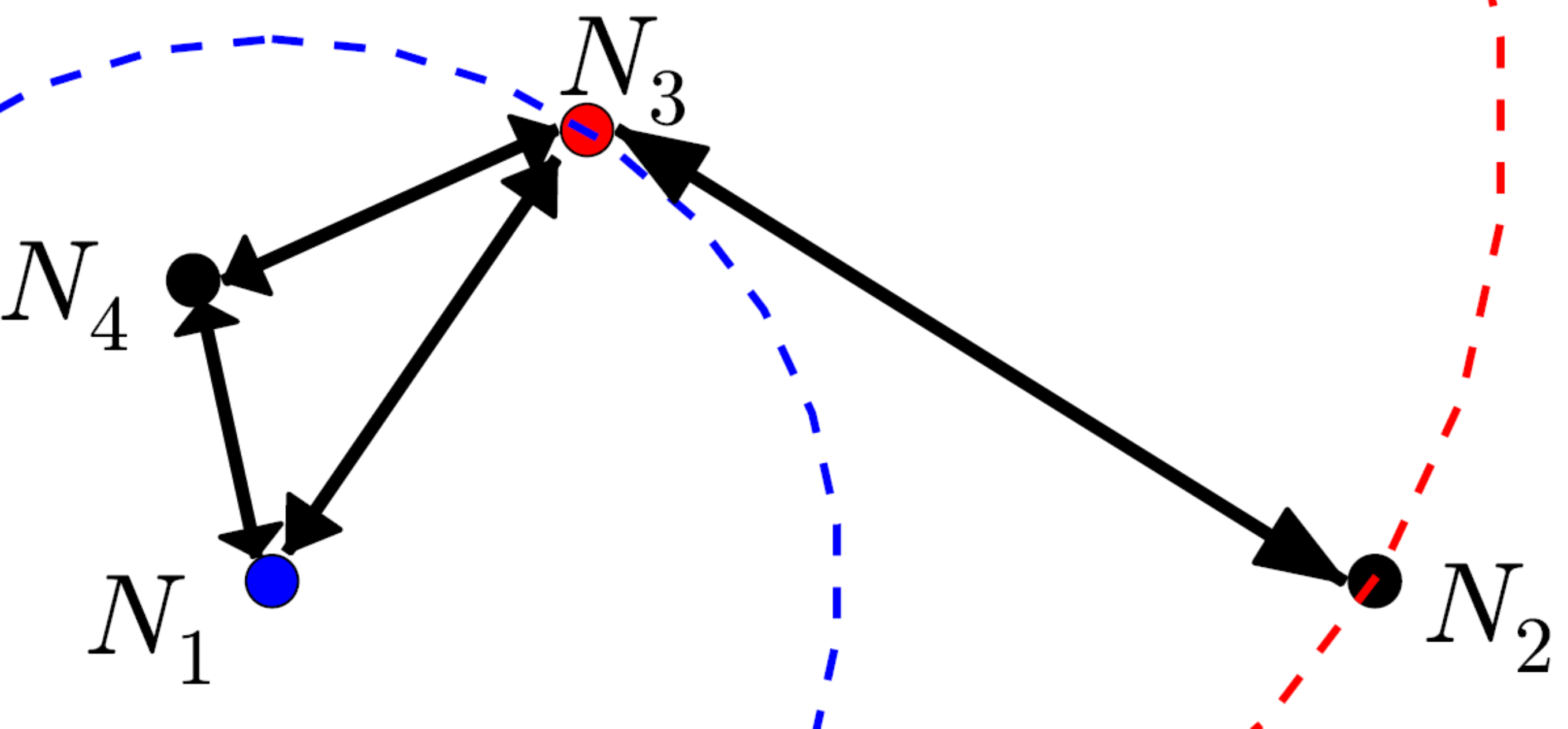}
        \label{fig:2-after-1-2}
        }\\
    \caption{An example illustrating how a node should choose its power level so as to increase the network's lifetime.}
    \label{fig:2-improve}
\end{center}
\vspace{-12pt}
\end{figure}
We further illustrate the definitions in this section using the topology shown
in Fig. \ref{fig:2-improve}. Suppose at time $t$, the topology of
the network is as shown in Fig. \ref{fig:2-before}. Denote by
$p(N_i,N_j)$ the minimum transmission power at which nodes $N_i$ and
$N_j$ have to transmit to reach each other. We refer to the set of
transmission powers that a node may switch to at time $t$ as its
\emph{available} transmission powers at time $t$. Then, according to
the topology given by
Fig. \ref{fig:2-before}, node $N_3$'s available transmission powers
are: $p(N_3,N_1)$, $p(N_3,N_2)$ and $p(N_3,N_4)$, while its current
transmission power is $p(N_3,N_1)$ (note that there is no need for
node $N_3$ to transmit at any power level other than the ones in this
{\em available} set).

Let $\mathbf{P}$ denote a mapping of nodes in the network to power
levels. For example, in Fig. \ref{fig:2-before}, the mapping
implemented is given by $\mathbf{P} = \{\, N_1 \rightarrow p(N_1,
N_2), N_2 \rightarrow p(N_2, N_3), N_3 \rightarrow p(N_3, N_1), N_4
\rightarrow p(N_4, N_3)\,\}$. Since node $N_3$ has the potential to
transmit at power $p(N_3,N_4)$ while still keeping the graph
connected, we refer to power $p(N_3,N_4)$ as node $N_3$'s
\emph{potential transmission power} under this node-to-power mapping
$\mathbf{P}$.

In general, the {\em potential transmission power} of a node is the
smallest available transmission power that the node can use such that
the graph is still connected while the power levels at all other nodes
remain the same. Let $p'_i(\mathbf{P})$ denote the potential transmission power of node $N_i$ under the
node-to-power mapping $\mathbf{P}$. Note that a node's potential
transmission power is no greater than its current transmission power
provided that the network is currently connected. That is,
$p'_i(\mathbf{P}) \leq p_i(t)$ if $G(t)$ is a connected graph and
$\mathbf{P}$ is the node-to-power mapping implemented at time $t$.

Transmitting at the potential transmission power as defined above can
increase the lifetime of a node beyond its estimated lifetime and,
consequently, of the network. To estimate the best lifetime a node can
achieve without changing the transmission powers of other nodes and
without disconnecting the network, we define the \emph{potential
  lifetime} of a node as the ratio between the node's current
remaining energy and its potential transmission power. Let
$L'_i(\mathbf{P},t)$ denote the potential lifetime of node $N_i$ at
time $t$ under the node-to-power mapping $\mathbf{P}$. Let
$p'_i(\mathbf{P})$ denote its potential transmission power under the
node-to-power mapping $\mathbf{P}$. Then, $L'_i(\mathbf{P},t) =
W_i(t)/p'_i(\mathbf{P})$.

If $p'_i(\mathbf{P}) \leq p_i(t)$, then $L_i(p_i(t),t)\leq
L'_i(\mathbf{P},t)$. Therefore, to increase its estimated lifetime,
a node should always try to change its current transmission power to
its potential transmission power, if they are not the
same. Figs. \ref{fig:2-before} and \ref{fig:2-after-2} illustrate
such a process for node $N_3$, where it changes its current
transmission power from $p(N_3,N_1)$ in Fig. \ref{fig:2-before} to
its potential transmission power $p(N_3,N_4)$ as illustrated in
Fig. \ref{fig:2-after-2}.

In Fig. \ref{fig:2-before}, suppose node $N_1$ is the node that has
the smallest estimated lifetime within the network. Then, $N_1$'s
estimated lifetime has to be improved in order to improve the
network's lifetime. In Fig. \ref{fig:2-after-2}, node $N_3$ reduces its
transmission power but this does not improve the potential
lifetime of node $N_1$. This implies that $N_1$'s estimated lifetime
cannot be improved by node $N_3$ reducing its transmission power. On
the other hand, if node $N_3$ chooses to transmit at a higher power
level, $p(N_3,N_2)$, as illustrated in Fig. \ref{fig:2-after-1}, then
node $N_1$'s potential transmission power reduces to
$p(N_1,N_3)$. Now, node $N_1$ can reduce its transmission power to the
new potential transmission power as illustrated in
Fig. \ref{fig:2-after-1-2}, thus improving the network's estimated
lifetime.

Let $R_i(t)$ denote the set of nodes $N_i$ can reach at time $t$; i.e.,
$R_i(t)=\{N_j\,|\,p(N_i,N_j)\leq p_i(t)\}$. Let $I_i(t)$ denote the set
of nodes that can reach node $N_i$ at time $t$, i.e,
$I_i(t)=\{N_j\,|\,p(N_j,N_i)\leq p_j(t)\}$. Then for any $N_j\in
I_i(t)$, we have $N_i\in R_j(t)$. We refer to the nodes in the set
$R_i(t)$ as the {\em reachable neighbors} of node $N_i$ and the nodes in the set
$I_i(t)$ as the {\em reverse-link neighbors} of $N_i$. For example, in
Fig. \ref{fig:2-before},
$N_1$'s reachable neighbors are $N_2$, $N_3$ and $N_4$ while $N_2$ has only one
reachable neighbor, $N_3$. Also, $N_1$ is a reverse-link neighbor of $N_2$,
$N_3$ and $N_4$. Let $H_i(t)=R_i(t)\cup N_i$ and let $O_i(t)=I_i(t)\cup N_i$.

In Fig. \ref{fig:2-before}, note that $N_1$'s potential  transmission
power will not be reduced unless either $N_3$ or $N_4$ has increased
its transmission power to be able to reach $N_2$. In general, $N_i$'s
reachable neighbors are the only nodes who can help $N_i$ reduce its potential
transmission power; and only the nodes who are
$N_i$'s reverse-link neighbors may benefit from $N_i$'s increase in
its transmission power level.

A question worth answering at this point is about what might be the
relationship between improving the network's lifetime and improving
node $N_1$'s estimated and potential lifetime. As we have stated
previously, the network's estimated lifetime is dependent upon
the node with the smallest estimated lifetime. If we compare the
power mapping illustrated in Fig. \ref{fig:2-before} and
Fig. \ref{fig:2-after-1-2}, we can see that node $N_3$ not only
has sacrificed its chance to improve its lifetime but also has
sacrificed its own estimated lifetime (increases its current
transmission power from $p(N_3,N_1)$ to $p(N_3,N_2)$) in order to
help improve node $N_1$'s potential lifetime which is the smallest in
Fig. \ref{fig:2-before}. When the topology is as illustrated in
Fig. \ref{fig:2-after-1-2}, suppose node $N_3$ is the node with the
smallest estimated lifetime. If its estimated lifetime in the
topology illustrated in Fig. \ref{fig:2-after-1-2} is less than
that of node $N_1$'s in the topology illustrated in
Fig. \ref{fig:2-before}, the network's estimated lifetime in fact
reduces rather than increases. In such a case, node $N_3$ should not
choose to increase its transmission power to help improve $N_1$'s
lifetime, but should instead focus on improving its own estimated
lifetime.

For any node $N_i$ within the graph, its estimated lifetime will not
improve unless it switches to its potential transmission power.
That is to say, no node can help node $N_i$ with its estimated
lifetime except for node $N_i$ itself; but as illustrated in
Fig. \ref{fig:2-improve}, another node may help with $N_i$'s
potential lifetime. In our example, node $N_3$ helps node $N_1$
improve its potential lifetime which eventually allows $N_1$ to
increase its estimated lifetime by changing its transmission power to
the potential transmission power.

The above discussion leads to the following primary and secondary
goals for each node in order to improve the network's lifetime while
also conserving energy as much as possible:
\begin{itemize}\label{goals}
\item {\em Primary goal:} Let $m(i)$ denote the node with the smallest
  potential lifetime amongst the reverse-link neighbors of node
  $N_i$. Let $q$ denote the potential lifetime of node $m(i)$. The
  primary goal of node $N_i$ should be to increase the potential
  lifetime of node $m(i)$ above $q$ while making sure that its own
  estimated lifetime does not reduce below $q$.
\item {\em Secondary goal:} The secondary goal of node $N_i$ should be
  to increase its own estimated lifetime.
\end{itemize}

The secondary goal is achieved once a node adopts its potential
transmission power as its current transmission power (in this
situation, its potential lifetime then becomes its estimated lifetime). Note
that, if the primary and the secondary goals conflict, a node should
always choose to meet the primary goal. This leads us into the design
of the cooperative game that each node can play with its reachable
neighbors and its reverse-link neighbors.

One may argue that since the ultimate goal is to extend functional
lifetime (the primary goal), the secondary goal should be
unnecessary. However, there are three reasons why we include the
secondary goal in our design of the algorithm. 

Firstly, for any realistic algorithm, it is not possible to precisely predict the 
network lifetime since it is not possible to predict with precision 
the future traffic load experienced by any given sensor node. As a result, an 
algorithm can only seek to maximize the {\em estimated} functional 
lifetime of a node and not the actual functional lifetime. Therefore, 
the secondary goal helps each sensor node save energy as much as 
possible so that when the estimated functional lifetime deviates from 
the actual functional lifetime, the node that actually determines the 
network's functional lifetime (the one that dies earliest) does not 
waste any energy by transmitting at a power larger than what was 
necessary to stay connected. 
Secondly, as discussed in Section \ref{sec:problemStatement}, we define
the functional lifetime assuming $c=0$. However, if the functional lifetime for some
application is defined assuming that the network function survives
until $c>0$ nodes are disconnected, a good heuristic for extending the
functional lifetime emerges if we implement the secondary goal in
conjunction with the primary goal assuming $c=0$ on the surviving
largest connected component every time a node dies. 
Thirdly, reducing the transmission power of a sensor node may help
reduce the interference among transmissions, reducing retransmissions
and also helping extend the functional lifetime of the network.


In the above discussion, without loss of generality, we assume that
there is only one node $m(i)$ with the smallest potential lifetime
amongst reverse-link neighbors of $N_i$. If there is a tie with two
nodes having identical potential lifetimes, one can always break the
tie in the algorithm using a consistently applied second criterion
such as the node id.

\section{The Ordinal Potential Game}\label{sec:game}

In this section, we present the notation and the utility function
governing the ordinal potential game into which we map the problem of
extending the lifetime of the network.

\subsection{Notation}

Table \ref{table:term} presents a glossary of terms used in this
section. In this paper, for brevity and clarity, we sometimes omit the
time index $t$ whenever the corresponding instant of time is clear
from the context.

\begin{table}[t]
\caption{A glossary of selected terms. If the time $t$ specified within each notation is clear within the
    context, it is omitted for purposes of brevity. In the glossary, however, we list each notation in both forms (with and without
    $t$).}
\begin{tabular}{lp{10.8cm}}
Notation & Definition \\
\hline  \\[-8pt]
$N$    &The set of $n$ nodes within the network. $N=\{N_1, N_2, . . . , N_n\}$.\\[1pt]
\hline \\[-8pt]
$G(t)$          & Graph representing the network topology at time $t$. $G(t)=(N, E(t))$\\[1pt]
\hline \\[-8pt]
$p_i(t)$ or $p_i$        & Node $N_i$'s transmission power at time
$t$. \\[1pt]
\hline \\[-8pt]
$\mathbf{P}$ & A mapping, $\mathbf{P} : N
\rightarrow \mathbb{R}$, of the nodes in the network to transmission
power levels. For example, the mapping actually implemented at time $t$ is
$\{ N_1 \rightarrow p_1(t), N_2 \rightarrow p_2(t),\dots \}$  \\[1pt]
\hline \\[-8pt]
$\mathbf{P}_{-i}$ &  A mapping of all the nodes except $N_i$ to
transmission power levels.  \\[1pt]
\hline \\[-8pt]
$p(N_i,N_j)$ & The minimum transmission power needed for node
$N_i$'s transmissions to reach node $N_j$. We assume $p(N_i,N_j)=p(N_j,N_i)$.\\[1pt]
\hline \\[-8pt]
$R_i(t)$ or $R_i$ & The set of nodes reachable by $N_i$'s transmissions at time
$t$. $R_i(t)=\{N_j\,|\,(N_i,N_j)\in E(t)\}$. This set of {\em reachable
neighbors} is also called the {\em reachable neighborhood} of $N_i$. \\[1pt]
\hline \\[-8pt]
$H_i(t)$ or $H_i$ & The set including $N_i$ and its reachable neighbors. $H_i(t)=R_i(t)\cup N_i$. \\[1pt]
\hline \\[-8pt]
$I_i(t)$ or $I_i$ & The set of nodes at time $t$ which can directly reach $N_i$
with their transmissions. $I_i(t)=\{N_j\,|\,(N_j,N_i)\in E(t)\}$. This set
of {\em reverse-link neighbors} of $N_i$ is also called {\em
  reverse-link neighborhood} of $N_i$.\\[1pt]
\hline \\[-8pt]
$O_i(t)$ or $O_i$ & The set including $N_i$ and its reverse-link
neighbors. $O_i(t)=I_i(t)\cup N_i$.
\\[1pt]
\hline \\[-8pt]
$\mathbf{P}_i$      & A mapping, $\mathbf{P}_i : H_i \rightarrow
\mathbb{R}$, of $N_i$ and its reachable neighbors to transmission
power levels. \\[1pt]
\hline \\[-8pt]
$\mathbf{P}_{i, -j}$      & A mapping of $N_i$ and its reachable
neighbors except $N_j$ to transmission power levels.\\[1pt]
\hline \\[-8pt]
$p'_i(\mathbf{P}_i)$ & $N_i$'s {\em potential} transmission power when
its and its reachable neighbors' power levels are given by the
mapping $\mathbf{P}_i$. \\[1pt]
\hline \\[-8pt]
$A_i$ & $N_i$'s possible transmission power choices.\\[1pt]
\hline \\[-8pt]
$W_i(t)$ or $W_i$& $N_i$'s remaining energy at time $t$. \\[1pt]
\hline \\[-8pt]
$L_i(a_i,t)$ & $N_i$'s estimated lifetime (remaining) at time $t$ if set to transmit at power $a_i$. $L_i(a_i,t)=W_i(t)/a_i$\\[1pt]
\hline \\[-8pt]
$L'_i(\mathbf{P}_i,t)$ & $N_i$'s {\em potential} lifetime at time $t$
when its and its reachable neighbors' transmission powers are given by
the mapping
$\mathbf{P}_i$. $L'_i(\mathbf{P}_i,t)=W_i(t)/p'_i(\mathbf{P}_i)$\\[1pt]
\hline \\[-8pt]
$m(i,t)$ or $m(i)$  &The node with the smallest potential lifetime
among $N_i$'s reverse-link neighbors at time $t$.
$m(i,t) = arg\min_{N_j\in I_i(t)} (L'_j(\mathbf{P}_j,t))$.\\[3pt]
\hline \\[-8pt]
\end{tabular}
\end{table}\label{table:term}

Suppose that at time $t=0$, each node is transmitting at its maximum
transmission power $p_{\text{max}}$. Then, the set of nodes that
includes $N_i$'s reachable neighbors is $R_i(0)=\{N_j\,|\,p(N_i,N_j)\leq
p_{\text{max}}\}$. Let $n_i$ denote the size of the set
$R_i(0)$. Therefore, we define the available transmission powers for node
$N_i$ as $A_i=\{p_i^1,p_i^2,\dots,p_i^{n_i}\}$ where, for any
$p_i^k\in A_i$, there exists a node $N_j\in R_i(0)$ such that $p_i^k$
is the minimum transmission power required for $N_i$ to reach
$N_j$. Note that a node does not need to transmit at power levels
other than the ones needed to reach other nodes within its maximum
range.

Let $\mathbf{P}$ denote a mapping of the nodes in the network to
transmission power levels. The mapping implemented at time
$t$ is $\mathbf{P} = \{\,N_i \rightarrow p_1(t), N_2 \rightarrow p_2(t), \dots \}$,
where $p_i(t)$ is the power at which node $N_i$ is set to make
transmissions at time $t$. We write $\mathbf{P}=\{\,f_i,\mathbf{P}_{-i}\,\}$ where
$f_i$ is a mapping of node $N_i$ to a certain power level and
$\mathbf{P}_{-i}$ is a mapping of all other nodes in the network to power levels.

Let $\mathbf{P}_i$ denote a mapping, $\mathbf{P}_i : H_i
\rightarrow \mathbb{R}$, of $N_i$ and its reachable neighbors to power
levels. We write $\mathbf{P}_i=\{\,f_j,\mathbf{P}_{i,-j}\,\}$ where
$f_j$ is a mapping of node $N_j \in H_i$ to a certain power level and
$\mathbf{P}_{i,-j}$ is a mapping of all other nodes in $H_i$ to power
levels.

Let $L_i(a_i, t)$ denote the estimated lifetime of node $N_i$ at time
$t$ if set to transmit at power level $a_i$. Per the definition of
estimated lifetime, $L_i(a_i, t)=W_i(t)/a_i$. If $a_i=p'_i(\mathbf{P}_i)$, then
$L_i(a_i,t)=L'_i(\mathbf{P}_i,t)$.

Denote by $m(i,t)$, or $m(i)$ for brevity, the node in $I_i(t)$ with the
smallest potential lifetime. We define the\emph{ potential lifetime of
$N_i$'s reverse-link neighborhood} at time $t$ as the potential
lifetime of node $m(i)$ at time $t$, i.e., $L'_{m(i)}(\mathbf{P}_{m(i)},t)$.

\subsection{The utility function} \label{sec:utility_function}

In the following, we present and justify the utility function
governing the ordinal potential game upon which our topology control
algorithm is based.

As stated in Section \ref{sec:analysis}, the primary goal for each
sensor node is to improve the potential lifetime of its reverse-link
neighborhood without also causing a reduction in the network's
estimated lifetime. While prioritizing the primary goal, the secondary
goal of the sensor node is to improve its own estimated lifetime. Both
of these goals are captured in the utility function presented in this
section.

Let $a_i\in A_i$ denote a power level that is available to node
$N_i$ at time $t$. Define the \emph{primary} utility function
(corresponding to the primary goal described in Section
\ref{sec:analysis}) for node $N_i$ with power level $a_i$ at time $t$ as:
\begin{align}\label{utility_primary}
u_i^X(a_i,t)&=\min\left( \min_{N_j\in
I_i(t)}L'_j(\{N_i \rightarrow
a_i,\mathbf{P}_{j,-i}\},t),L_i(a_i,t)\right) \nonumber\\
&=\min\left(L'_{m(i)}(\{N_i \rightarrow a_i,\mathbf{P}_{m(i),-i}\},t),L_i(a_i,t)\right)
\end{align}
This is the minimum of the estimated lifetime of node $N_i$ at power
level $a_i$ and the potential lifetime of the node whose value is the
minimum amongst its reverse-link neighbors. Maximizing this is the
primary goal as explained in Section \ref{sec:analysis}.

Define the \emph{secondary} utility function for node $N_i$
with power level $a_i$ at time $t$ as:
\begin{equation}\label{utility_secondary}
u_i^Y(a_i,t)=L_i(a_i,t)
\end{equation}
This is the estimated lifetime of node $N_i$ when transmitting at
power level $a_i$ at time $t$. Maximizing this is the secondary goal
as also explained in Section \ref{sec:analysis}.

Bearing in mind the two goals, primary and secondary, for each node,
we define the utility function $u_i$ for node $N_i$ with power level
$a_i$ at time $t$ as:
\begin{align}\label{utility_simplified}
u_i(a_i,t)&=c_i(a_i,t) \left[ u_i^X(a_i,t)+\ell_i(a_i,t) u_i^Y(a_i,t)\right]\nonumber\\
&=c_i(a_i,t)\min\left(L'_{m(i)}(\{N_i \rightarrow a_i,,\mathbf{P}_{m(i),-i}\},t),L_i(a_i,t)\right)+ c_i(a_i,t)\ell_i(a_i,t)L_i(a_i,t)
\end{align}
where $c_i(a_i,t)$ and $\ell_i(a_i,t)$ are defined in the following
paragraphs.

The term $c_i(a_i,t)$ in Eqn. (\ref{utility_simplified}) is a binary
function indicating whether node $N_i$, when set to transmit at power
$a_i$, is connected to every node $N_j\in R_i(t)$.
More specifically,
\[
  c_i(a_i,t)=\left\{
   \begin{aligned}
   &1, \text{\ if a path exists from $N_i$ to each $N_j\in R_i(t)$}\\
   &0, \text{\ otherwise}
   \end{aligned}
   \right.\nonumber
\]
If node $N_i$ has lost connectivity with a certain node $N_j$
by transmitting at power $a_i$ at time $t$, i.e., $c_i(a_i,t)=0$,
then, the network's connectivity is lost and by Case 1 in Section
\ref{sec:problemStatement}, the network's life has ended. This should
be reflected in node $N_i$'s own utility function, and thus,
$u_i(a_i,t)=0$ when $c_i(a_i,t)=0$. 
Note that checking for the existence of a path to every $N_j\in
R_i(t)$ is a localized function and does not require global or
centralized knowledge.


The term $\ell_i(a_i,t)$ in Eqn. (\ref{utility_simplified}) is a
binary function indicating whether the node's own estimated lifetime
should be considered when calculating its utility at power level
$a_i$. In the following, we will describe the conditions under which
$\ell_i(a_i,t)$ takes on the values of either $0$ or $1$.

As discussed in Section \ref{sec:analysis}, improving its own lifetime
is only the secondary goal for every sensor node. When the primary and
the secondary goals of a node are in conflict, the secondary goal of
improving its own lifetime should yield to the primary
goal. Therefore, for $a_i$ that leads to this situation,
$\ell_i(a_i,t) = 0$ indicating that the secondary goal of node $N_i$
yields to the primary goal. On the other hand, for the power level
$a_i$ at which node $N_i$ is able to achieve its primary goal without
conflict with the secondary goal, $\ell_i(a_i,t)$ should take on the
value of 1 and node $N_i$ should now focus on its secondary goal as
well. In cases where node $N_i$ is not able to achieve the primary
goal at whichever power it is transmitting, it should then focus on
improving its own estimated lifetime and therefore, the function
$\ell_i(a_i,t)$ then takes on the value of 1 for every $a_i$
selected. The following paragraphs present a formal definition of
function $\ell_i(a_i,t)$.

Suppose at power level $a_i$, node $N_i$ is able to help node $m(i,t)$ reduce its potential
transmission power, and its own estimated lifetime at power $a_i$ is
larger than its reverse-link neighborhood's previous potential
lifetime. Under such a circumstance, we refer to power level $a_i$ as
a \emph{preferred power level} of node
$N_i$. Note that, for a node-to-power mapping $\mathbf{P}$, there may exist
several preferred power levels for node $N_i$. We denote the set of
preferred power levels for node $N_i$ under the node-to-power
mapping $\mathbf{P}$
as $K_i(\mathbf{P_{-i}})$. For any power level $a_i\in
K_i(\mathbf{P_{-i}})$, node $N_i$'s reverse-link neighborhood's potential
lifetime is extended by node $N_i$ transmitting at power $a_i$ and
node $N_i$'s lifetime at power level $a_i$ exceeds its previous
reverse-link neighborhood's potential lifetime. Therefore, at a power level $a_i\in
K_i(\mathbf{P_{-i}})$, the primary goal for node $N_i$ is met and node $N_i$
should focus on optimizing for its own estimated lifetime (the
secondary goal) through the utility function. Therefore, we can
conclude that $\ell_i(a_i,t)=1$ in such a case.

If there exists no preferred power level $a_i$ at which $N_i$ can
transmit to increase its primary utility function (indicated by
$K_i(\mathbf{P_{-i}})=\emptyset$), then also node $N_i$ should focus
on improving its own estimated lifetime through the utility
function. In this case, $N_i$'s lifetime should be still be relevant
in the local utility function $u_i(a_i, t)$ and so, $\ell_i(a_i,t)=1$
for any value of $a_i$.

If neither of the above two cases is valid, then transmitting at power
level $a_i$ causes a conflict between node $N_i$'s primary and
secondary goals. In this case, node $N_i$ should focus on its primary
goal only and therefore, $\ell_i(a_i,t)=0$.

Based on the above reasoning, $\ell_i(a_i,t)$ is defined as:
\begin{equation}
  \ell_i(a_i,t)=\left\{
   \begin{aligned}
   &1, \text{\ \ } \mathrm{if~} K_i(\mathbf{P_{-i}}) = \emptyset, \mathrm{~or~}
   a_i \in K_i(\mathbf{P_{-i}})\\
   &0, \text{\ \ otherwise}   \\
   \end{aligned}
   \right.\nonumber
\end{equation}

\subsection{The ordinal potential game}

We are now ready to describe the strategic game $\Gamma=\langle
N,A,U\rangle$ as having the following three components:
\begin{itemize}
\item Player set $N$: $N_i\in N = \{N_1, N_2, . . . , N_n\}$ where $n$ is the number of
nodes in the network.
\item Action set $A$: $a\in A= \Pi_{i=1}^n A_i$ is the space of all action vectors,
where each component $A_i$ represents the set of available power
levels at which $N_i$ may transmit.
\item Utility function set $U$: For each player $N_i$, utility
function $u_i: A\rightarrow \mathbb{R}$ as given by
Eqn. (\ref{utility_simplified}) which models the node's preferences
for its available power level choices. The vector of these utility
functions is $U: A\rightarrow \mathbb{R}^n$.
\end{itemize}

\begin{theorem}
The game $\Gamma=\langle N,A,U\rangle$ is an ordinal potential game
and its ordinal potential function is given by
\begin{equation}
\Phi(\mathbf{P},t)=C(\mathbf{P})\min_{N_i\in
N}L'_i(\mathbf{P}_i,t)\label{potential}
\end{equation}
where $C(\mathbf{P})$ is the binary connectivity function indicating
whether the graph is connected with node-to-power mapping $\mathbf{P}$, i.e,
\begin{equation}
  C(\mathbf{P})=\left\{
   \begin{aligned}
   &1, \text{\ \ if the graph is connected }\\
   &0, \text{\ \ otherwise}   \\
   \end{aligned}
   \right.\nonumber
\end{equation}
\begin{proof}
We will prove this by applying the definition of an ordinal potential
game and proving that, at any time instant $t$, the difference in individual utilities for each
node from unilaterally changing its strategy and the difference in
values of the global potential function have the same sign \cite{MonSha1996,MacDaS2006}. Denote the
mapping of nodes to power levels as follows: when $N_i$ is
transmitting at power level $a_i$ as $\mathbf{P}=\{ N_i \rightarrow
a_i,\mathbf{P}_{-i}\}$ and when $N_i$ is transmitting at power level
$a_i'$ as $\mathbf{P}'=\{ N_i \rightarrow
a_i',\mathbf{P}_{-i}\}$. First, for the difference in an individual
node's utilities,
we have:
\begin{equation}
\Delta u_i(t)=u_i(\mathbf{P},t)-u_i(\mathbf{P}',t) \nonumber
\end{equation}
Omitting $t$ for brevity, we can rewrite this equation as:
\begin{align}
\Delta u_i&=u_i(\mathbf{P})-u_i(\mathbf{P}') \nonumber\\
&\hskip-.2em=c_i(a_i)L_i(a_i)\ell_i(a_i)+c_i(a_i)\min \left\{
    L'_{m(i)}(\mathbf{P}),L_i(a_i) \right\} \nonumber\\
&\hskip-.2em-c_i(a_i')L_i(a_i')\ell_i(a_i')-c_i(a_i')\min\left\{ L'_{m(i)}(\mathbf{P'}),L_i(a_i')\right\}\nonumber
\end{align}
Note that, with $N_i$'s power level being either $a_i$ or $a_i'$,
the power levels for the rest of the nodes within the network remain
the same. Since at any time instant $t$, $N_j\in O_i(t)$ are the only nodes whose potential
lifetime may be affected by $N_i$'s power level, we can thus
conclude that for node $N_j\not\in O_i(t)$,
$L'_j(\mathbf{P},t)=L'_j(\mathbf{P}',t)$.

Now, the difference in the values of the global potential function,
$\Delta\Phi(t)$, is:
\[\Delta\Phi(t)=C(\mathbf{P})\min_{N_i\in
N}L'_i(\mathbf{P},t)-C(\mathbf{P'})\min_{N_i\in
N}L'_i(\mathbf{P'},t)\]
Since this equation holds for any value of $t$, we can omit $t$ to
simplify the notation.
\begin{align}
\Delta\Phi & =C(\mathbf{P})\min_{N_i\in N}L'_i(\mathbf{P})-C(\mathbf{P'})\min_{N_i\in
N}L'_i(\mathbf{P'}) \nonumber\\
& =C(\mathbf{P})\min\left\{ \min_{N_j\in
O_i}L'_j(\mathbf{P}),\min_{N_k\not\in{O_i}}L'_k(\mathbf{P}))\right\}-C(\mathbf{P'})\min\left\{ \min_{N_j\in
O_i}L'_j(\mathbf{P'}),\min_{N_k\not\in{O_i}}L'_k(\mathbf{P'})\right\}\nonumber\\
&=C(\mathbf{P})\min\left\{T_i(\mathbf{P}),T_{-i}(\mathbf{P})\right\} -C(\mathbf{P}')\min\left\{T_i(\mathbf{P}'),T_{-i}(\mathbf{P}')\right\}\nonumber
\end{align}
where $T_i(\mathbf{P})=\min_{N_j\in O_i}L'_j(\mathbf{P})$ is the
smallest potential lifetime amongst node $N_i$ and its reverse-link
neighborhood when the node-to-power mapping is $\mathbf{P}$. Recall
that $O_i = N_i \cup I_i$ and therefore:
\begin{eqnarray}
T_i(\mathbf{P})&=&\min \{ L'_i(\mathbf{P}), \min_{N_j\in I_i}L'_j(\mathbf{P})\}\nonumber\\
&=&\min\{L'_i(\mathbf{P}), L'_{m(i)}(\mathbf{P})\}
\end{eqnarray}

$T_i(\mathbf{P}')$ is similarly defined. $T_{-i}(\mathbf{P})$ and $T_{-i}(\mathbf{P'})$ are
also defined similarly as $\min_{N_k\not\in{O_i}}L'_k(\mathbf{P})$ and
$\min_{N_k\not\in{O_i}}L'_k(\mathbf{P'})$, respectively. Since nodes
within $O_i$ are the only nodes whose potential lifetime may be
influenced by node $N_i$'s change in its transmission power, we can
therefore conclude that $T_{-i}(\mathbf{P})=T_{-i}(\mathbf{P'})$.

Without loss of generality, we assume that $a_i>a_i'$, indicating
that if $C(\mathbf{P}')=1$, then $C(\mathbf{P})=1$. We can also
conclude that $L_i(a_i)<L_i(a_i')$. According to the definition
of $C(\mathbf{P})$, if $a_i=0$, then $C(\mathbf{P})=0$. The possible
cases of $c_i(a_i,t)$ and $c_i(a_i',t)$ are (omitting $t$ for brevity):
\begin{itemize}
\item Case 1: $c_i(a_i)=c_i(a_i')=0 \Rightarrow C(\mathbf{P})=C(\mathbf{P}')=0$
\item Case 2: $c_i(a_i)=1, c_i(a_i')=0 \Rightarrow C(\mathbf{P}')=0$
\item Case 3: $c_i(a_i)=c_i(a_i')=1$
\end{itemize}
In Cases 1 and 2, the network is not connected with $a_i$ or
$a_i^\prime$ or both. In these cases, it is easy to prove that $\Delta
u_i$ and $\Delta \Phi$ have the same sign. We consider Case 3 in
detail in the following.

In Case 3, the local graph within $N_i$'s range is connected whether
$N_i$'s power level is $a_i$ or $a_i'$. Since all other
nodes except $N_i$'s power levels remain the same at time $t$,
we can conclude that $C(\mathbf{P})=C(\mathbf{P'})$. This leads us
to two situations: in one, $C(\mathbf{P})=C(\mathbf{P'})=0$, i.e, the
full graph $G$ is not connected because of some node located outside
of $N_i$'s range, and in the other, $C(\mathbf{P})=C(\mathbf{P'})=1$,
i.e, the full graph $G$ is connected. In the case the graph is not
connected, $C(\mathbf{P})=C(\mathbf{P'})=0$ and, therefore, $\Delta
\Phi_i=0$. Thus, we can conclude that $\Delta u_i$ and $\Delta
\Phi_i$ have the same sign. In the following, we now focus on the
situation in which the full graph $G$ is connected.

The Case 3 situation in which the graph is connected, i.e.,
$C(\mathbf{P})=C(\mathbf{P'})=1$, can be further categorized into four
sub-cases:
\begin{itemize}
\item Sub-case (3a):\\
\mbox{~~~}$\min\{ T_i(\mathbf{P}),T_{-i}(\mathbf{P})\}= T_i(\mathbf{P})$, and $\min\{T_i(\mathbf{P'}), T_{-i}(\mathbf{P'})\}=T_i(\mathbf{P'})$.
\item Sub-case (3b):\\
\mbox{~~~}$\min\{ T_i(\mathbf{P}),T_{-i}(\mathbf{P})\}=
T_{-i}(\mathbf{P})$, and $\min\{T_i(\mathbf{P'}),
T_{-i}(\mathbf{P'})\}=T_i(\mathbf{P'})$.
\item Sub-case (3c):\\
\mbox{~~~}$\min\{ T_i(\mathbf{P}),T_{-i}(\mathbf{P})\}=
T_i(\mathbf{P})$, and $\min\{T_i(\mathbf{P'}), T_{-i}(\mathbf{P'})\}= T_{-i}(\mathbf{P'})$.
\item Sub-case (3d):\\
\mbox{~~~}$\min\{
T_i(\mathbf{P}),T_{-i}(\mathbf{P})\}=T_{-i}(\mathbf{P})$, and $\min\{T_i(\mathbf{P'}), T_{-i}(\mathbf{P'})\}=T_{-i}(\mathbf{P'})$.
\end{itemize}

{\em Case (3a):} In this case, whether $N_i$'s power level is $a_i$ or $a_i'$, the
node with the smallest potential lifetime lies either within $N_i$'s
reverse-link neighborhood or is node $N_i$ itself. Since
$T_i(\mathbf{P})=\min\{L'_{m(i)}(\mathbf{P}),L'_i(\mathbf{P})\}$ and
$T_i(\mathbf{P'})=\min\{L'_{m(i)}(\mathbf{P'}),L'_i(\mathbf{P'})\}$,
we can conclude that:

\begin{align}
\Delta\Phi&=T_i(\mathbf{P})-T_i(\mathbf{P'})=\min\left\{L'_{m(i)}(\mathbf{P}),L'_i(\mathbf{P})\right\}-\min\left\{L'_{m(i)}(\mathbf{P'}),L'_i(\mathbf{P'})\right\} \label{eqn:Phi}\\
\Delta u_i&=L_i(a_i)\times\ell_i(a_i)-L_i(a_i')\times\ell_i(a_i')+\min\left\{L'_{m(i)}(\mathbf{P}),L_i(a_i)\right\}-\min\left\{L'_{m(i)}(\mathbf{P'}),L_i(a_i')\right\}\label{eqn:u}
\end{align}

Since a node's potential transmission power is no larger than its
current transmission power, we can conclude that $p_i(\mathbf{P})\leq
a_i$, and $p_i(\mathbf{P'})\leq a_i'$. Also, $a_i$ is at least one
power level larger than $a_i'$ and, therefore, $p_i(\mathbf{P'})\leq
a_i'\leq p_i(\mathbf{P})\leq a_i$. We conclude:
\begin{equation}
L'_i(\mathbf{P'})\geq L_i(a_i')\geq L'_i(\mathbf{P})\geq L_i(a_i)\label{eqn:relationship}
\end{equation}

Now, there are four sub-sub-cases based on the values of $\ell_i(a_i)$
and $\ell_i(a_i')$:
\begin{itemize}
\item Case (3a-i): $\ell_i(a_i)=\ell_i(a_i')=1$
\item Case (3a-ii): $\ell_i(a_i)=\ell_i(a_i')=0$
\item Case (3a-iii): $\ell_i(a_i)=1$, and $\ell_i(a_i')=0$
\item Case (3a-iv): $\ell_i(a_i)=0$, and $\ell_i(a_i')=1$
\end{itemize}
In the following, we consider each of the above sub-sub-cases.

{\em Case (3a-i):} According to the definition of $\ell_i(a_i,t)$,
either both power levels $a_i$ and $a_i'$ can help improve node
$N_i$'s reverse-link neighborhood's potential lifetime or neither of
them can. Therefore, we have
$L'_{m(i)}(\mathbf{P})=L'_{m(i)}(\mathbf{P'})$. Note that,
$L'_i(\mathbf{P})\leq L'_i(\mathbf{P'})$. Now, we can rewrite Eqn. (\ref{eqn:Phi}) as
\begin{align}
\Delta\Phi&=\min\left\{L'_{m(i)}(\mathbf{P}),L'_i(\mathbf{P})\right\}-\min\left\{L'_{m(i)}(\mathbf{P'}),L'_i(\mathbf{P'})\right\}\nonumber\\
&=\min\left\{L'_{m(i)}(\mathbf{P'}),L'_i(\mathbf{P})\right\}-\min\left\{L'_{m(i)}(\mathbf{P'}),L'_i(\mathbf{P'})\right\}\nonumber\\
&\leq \min\left\{L'_{m(i)}(\mathbf{P'}),L'_i(\mathbf{P'})\right\}-\min\left\{L'_{m(i)}(\mathbf{P'}),L'_i(\mathbf{P'})\right\}\nonumber\\
&=0\nonumber
\end{align}
As for $\Delta u_i$, since $L_i(a_i)< L_i(a_i')$, and
$L'_{m(i)}(\mathbf{P})=L'_{m(i)}(\mathbf{P'})$, we can rewrite Eqn. (\ref{eqn:u}) as:
\begin{align}
\Delta u_i&=L_i(a_i)-L_i(a_i')+\min\left\{L'_{m(i)}(\mathbf{P}),L_i(a_i)\right\}-\min\left\{L'_{m(i)}(\mathbf{P'}),L_i(a_i')\right\}\nonumber\\
&<\min\left\{L'_{m(i)}(\mathbf{P'}),L_i(a_i)\right\}-\min\left\{L'_{m(i)}(\mathbf{P'}),L_i(a_i')\right\}\nonumber\\
&\leq \min\left\{L'_{m(i)}(\mathbf{P'}),L_i(a_i')\right\}-\min\left\{L'_{m(i)}(\mathbf{P'}),L_i(a_i')\right\}\nonumber\\
&=0\nonumber
\end{align}
Therefore, we have $\Delta\Phi\leq 0$ and $\Delta u_i<0$. Thus, as for
Case (3a-i), $\Delta\Phi$ and $\Delta u_i$ share the same sign.

{\em Case (3a-ii):} Since $\ell_i(a_i)=\ell_i(a_i')=0$, it
indicates that node $N_i$'s reverse-link neighborhood's potential
lifetime cannot be extended when node $N_i$ is transmitting at either
power level $a_i$ or $a_i'$. Thus, we can conclude that
$L'_{m(i)}(\mathbf{P})=L'_{m(i)}(\mathbf{P'})$. Following a similar
line of deduction as in Case (3a-i), we can conclude that
$\Delta\Phi\leq 0$.

As for $\Delta u_i$, also following the same line of deduction as in
Case (3a-i), we have:
\begin{align}
\Delta u_i&=\min\left\{L'_{m(i)}(\mathbf{P}),L_i(a_i)\right\}-\min\left\{L'_{m(i)}(\mathbf{P'}),L_i(a_i')\right\}\nonumber\\
&\leq\min\left\{L'_{m(i)}(\mathbf{P'}),L_i(a_i)\right\}-\min\left\{L'_{m(i)}(\mathbf{P'}),L_i(a_i')\right\}\nonumber\\
&=0\nonumber
\end{align}
This implies that both $\Delta\Phi$ and $\Delta u_i$ are no larger
than 0 and, therefore, share the same sign.

{\em Case (3a-iii):} The fact that $\ell_i(a_i)=1$ and
$\ell_i(a_i')=0$ indicates that node $N_i$'s preferred power set is
not empty and power level $a_i$ is one of the preferred power levels
while power level $a_i'$ is not. Therefore, by node $N_i$ transmitting
at power $a_i$, its reverse-link neighborhood's potential lifetime can
be extended. Denote node $N_i$'s reverse-link neighborhood's previous
potential lifetime by $L'_{\text{pre}}$. We know that
$L'_{m(i)}(\mathbf{P})>L'_{\text{pre}}$. On the other hand, since
$\ell_i(a_i')=0$ and $a_i>a_i'$, we can conclude that by
node $N_i$ transmitting at power level $a_i'$, its reverse-link
neighborhood's potential lifetime cannot be improved. Thus, we have
$L'_{m(i)}(\mathbf{P'})=L'_{\text{pre}}<L'_{m(i)}(\mathbf{P})$. Also,
according to the definition of $\ell_i(a_i)$, we can conclude
$L_i(a_i)>L'_{\text{pre}}$. Together with
Eqn. (\ref{eqn:relationship}), we can conclude that
$L'_i(\mathbf{P})\geq L_i(a_i)>L'_{m(i)}(\mathbf{P'})$. Thus, we can
rewrite Eqns. (\ref{eqn:Phi}) and (\ref{eqn:u}) as:
\begin{align}
\Delta\Phi&=\min\left\{L'_{m(i)}(\mathbf{P}),L'_i(\mathbf{P})\right\}-\min\left\{L'_{m(i)}(\mathbf{P'}),L'_i(\mathbf{P'})\right\}\nonumber\\
&\geq\min\left\{L'_{m(i)}(\mathbf{P'}),L'_i(\mathbf{P})\right\}-L'_{m(i)}(\mathbf{P'})\nonumber\\
&>\min\left\{L'_{m(i)}(\mathbf{P'}),L'_{m(i)}(\mathbf{P'})\right\}-L'_{m(i)}(\mathbf{P'})=0\nonumber\\
\Delta u_i&=L_i(a_i)-\min\left\{L'_{m(i)}(\mathbf{P'}),L_i(a_i')\right\}+\min\left\{L'_{m(i)}(\mathbf{P}),L_i(a_i)\right\}\nonumber\\
&\geq L_i(a_i)-L'_{m(i)}(\mathbf{P'})+\min\left\{L'_{m(i)}(\mathbf{P'}),L_i(a_i)\right\}\nonumber\\
&>\min\left\{L'_{m(i)}(\mathbf{P'}),L_i(a_i)\right\}>0\nonumber
\end{align}
Therefore, we have proved that, in Case (3a-iii), both $\Delta\Phi$
and $\Delta u_i$ are positive numbers, and therefore, share the same
sign.

{\em Case (3a-iv):} The fact that $\ell_i(a_i)=0$ indicates that the
preferred power level set $K_i(\mathbf{P_{-i}})$ is not empty
and power level $a_i$ is not within $K_i(\mathbf{P_{-i}})$. On the
other hand, since $\ell_i(a_i')=1$, and
$\mathbf{P_{-i}}=\mathbf{P'_{-i}}$, we can conclude that
$a_i'$ is a preferred power level and
$L'_{\text{pre}}<L_i(a_i')$. This also indicates that when
transmitting at power level $a_i'$, node $N_i$ serves as a relay node
for node $m(i)$ enabling it to reduce its transmission power without
disconnecting the network. Therefore, by transmitting at power
$a_i>a_i'$, node $N_i$ should also be able to serve as the bridge node
for node $m(i)$. Thus, we can conclude that
$L'_{m(i)}(\mathbf{P})=L'_{m(i)}(\mathbf{P'})$. Following similar
lines of deduction as in Cases (3a-i) and (3a-ii), we conclude that
$\Delta\Phi\leq 0$.

Now, since node $m(i)$'s potential lifetime can be improved by node
$N_i$ transmitting at power level $a_i$, therefore, the only reason
why $\ell_i(a_i)=0$ is that by transmitting at this power, node
$N_i$'s lifetime at power level $a_i$ is less than node $m(i)$'s
previous potential lifetime, i.e.,
$L_i(a_i)<L'_{\text{pre}}<L_i(a_i')$. We therefore can rewrite Eqn. \ref{eqn:u} as:
\begin{align}
\Delta
u_i&=\min\left\{L'_{m(i)}(\mathbf{P}),L_i(a_i)\right\}-\min\left\{L'_{m(i)}(\mathbf{P'}),L_i(a_i')\right\}-L_i(a_i')\nonumber\\
&\leq \min\left\{L'_{m(i)}(\mathbf{P'}),L_i(a_i')\right\}-\min\left\{L'_{m(i)}(\mathbf{P'}),L_i(a_i')\right\}-L_i(a_i')\nonumber\\
&=0-L_i(a_i')<0\nonumber
\end{align}
Therefore, we have proved that, in Case (3a-iv), $\Delta\Phi$ and $\Delta u_i$ share the same sign.

From the above arguments, we have proved that $\Delta \Phi_i$ and
$\Delta u_i$ hold the same sign for all possible sub-cases in Case (3a).

{\em Case (3b):} We have
$T_i(\mathbf{P'})<T_{-i}(\mathbf{P'})=T_{-i}(\mathbf{P})<T_i(\mathbf{P})$. This
indicates that by node $N_i$ transmitting at power $a_i$, its primary
goal has been met. Therefore, we have $\ell_i(a_i)=1$, and
$\ell_i(a_i')=0$. Then, we have $\Delta
\Phi_i=T_{-i}(\mathbf{P})-T_i(\mathbf{P'})>0$ and $\Delta
u_i=L_i(a_i)+T_i(\mathbf{P})-T_i(\mathbf{P'})>L_i(a_i)>0$. Therefore,
in Case (3b), $\Delta \Phi_i$ and $\Delta u_i$ hold the same sign.

{\em Case (3c):} In this case,
$T_i(\mathbf{P})<T_{-i}(\mathbf{P})=T_{-i}(\mathbf{P'})<T_i(\mathbf{P'})$. Therefore,
we have $\Delta \Phi_i=T_i(\mathbf{P})-T_{-i}(\mathbf{P'})<0$. Exactly
as in Case (3a), there are four sub-cases depending on the values of
$\ell_i(a_i)$ and $\ell_i(a_i')$. For Cases (3c-i), (3c-ii) and
(3c-iv), we can follow similar lines of deduction as in Cases
(3a-i), (3a-ii) and (3a-iv) to prove that $\Delta u_i\leq 0$ and,
therefore, $\Delta \Phi_i$ and $\Delta u_i$ hold the same sign.

As for Case (3c-iii), it can be shown that it is impossible. Following
the logic discussed in Case (3a-iii), we have
$\Phi_i=T_i(\mathbf{P})-T_{-i}(\mathbf{P'})>T_i(\mathbf{P})-T_i(\mathbf{P'})>0$,
which is in contradiction to the assumption in Case (3c-iii)
that $T_i(\mathbf{P})<T_i(\mathbf{P'})$.

So, in all sub-cases of Case (3c), $\Delta \Phi_i$ and $\Delta u_i$
have the same sign.

{\em Case (3d):} In this case, we can conclude that $\Delta
\Phi_i=T_{-i}(\mathbf{P})-T_i(\mathbf{P'})=0$. Therefore, no matter
what the sign of $\Delta u_i$, we can conclude that $\Delta \Phi_i$
and $\Delta u_i$ have the same sign.

This concludes the consideration of all possible cases and sub-cases,
in all of which we have shown that $\Delta \Phi(\mathbf{P})$ and
$\Delta u_i(\mathbf{P})$ have the same sign. This proves that
$\Phi(\mathbf{P},t)$ is an ordinal potential function of
$u_i(\mathbf{P},t)$, and $\Gamma$ is an ordinal potential game.
\end{proof}
\end{theorem}

Since this is an ordinal potential game, seeking the optimal global
potential function yields a Nash equilibrium \cite{MonSha1996,MacDaS2006}. In the next section, we
propose a distributed localized algorithm that adaptively seeks to
optimize the global potential function through each node seeking to
optimize its own utility function defined in
Eqn. (\ref{utility_simplified}).

\section{The CTCA algorithm}\label{sec:pseudo_code}
This section presents the Cooperative Topology
Control with Adaption (CTCA) algorithm in which each node plays the
ordinal potential game, discussed in the previous section, with the goal
of increasing network lifetime.

\subsection{Pseudo-code and rationale}

We use the same terminology as in the previous section, but for
brevity, we omit the time $t$ in our notation.
\begin{algorithm}[t]
\KwIn{Maximum transmission power $P_{\text{max}}$}
\KwOut{$G_i'=(V_i,E_i')$, the local topology of node $N_i$}
\textbf{Initialization phase:}\\
 Broadcast a Hello message with power $p_{\text{max}}$\;
 Compile $R_i$\;
 $k\leftarrow $ number of reachable neighbors in $R_i$\;
 Compile $A_i=\{p(N_i,N_j)\,|\,N_j\in R_i\}$\;
 Sort $A_i$ such that $A_i[1]<A_i[2]< \dots < A_i[k]$\;
 Broadcast neighbor info $(N_j,p(N_i,N_j))$ for $N_j\in R_i$ with power $A_i[k]$\;
 Receive the information sent by neighbor $N_j\in R_i$\;
 Run DLSS algorithm, determine $p_i$, compile $R_i$\;
 Broadcast $p_i$ with power level $A_i[k]$\;
 Receive $p_j$ from $N_j\in R_i$, and compile $I_i$\;
 $S_i\leftarrow$ {\em AbleToReducePower}$(N_i,p_i)$\;
 Broadcast $S_i$ with power $p_i$\;
 Receive $S_j$ from $N_j\in I_i$\;
\BlankLine
 \textbf{Power adjustment phase:}\\
 {\em EnergyInfoShared} $\leftarrow$ False\;
\While {$W_i>0$}
{ $q\leftarrow 0$\;
\If{not {\em EnergyInfoShared}}
 {Broadcast $W_i$ with power $p_i$\;
 {\em EnergyInfoShared} $\leftarrow$ True\;
}
 Send remaining energy request for $N_j\in I_i$\;
 Receive $W_j$ from $N_j\in I_i$\;
 Wait for a random time $t\in [0,T_1]$\;
 $p_i,q\leftarrow$ NAPA($p_i,q$)\;
 Wait for $T_2$ time\;
 {\em EnergyInfoShared} $\leftarrow$ False\;
 Wait for $T_3-T_2$ time\;
}
\textbf{return} $G_i$
\caption{CTCA algorithm executed at node $N_i$}\label{TC:CTCA}
\end{algorithm}
The pseudo-code of the CTCA algorithm is presented in
Fig.\,\,\ref{TC:CTCA}. The initialization phase of the algorithm
(lines 01--13) enables each node to rapidly reduce its transmission
power (using the DLSS algorithm executed in line 08), compile $A_i$ (the
list of power levels that $N_i$ can switch to) and prepare for the
power adjustment phase illustrated in lines 14--29.

To offer a dynamic environment where each node updates its
transmission power periodically, the algorithm operates in rounds. At
the beginning of each round, each node broadcasts its current remaining energy if it has not been
broadcasted before, which is indicated by the {\em EnergyInfoShared}
flag. This process is described in lines 17--19. It will also send out a request
for its reverse-link neighbors' current remaining energy levels and
update $W_j$ for $N_j\in I_i$ based on the received data as described
in lines 21--22. The node will then wait for a random period of time
$t$ ranging from $0$ to $T_1$ before executing the {\em
Neighbor-Assisted Power Adjust (NAPA)} function to adjust its
transmission power. NAPA is the game-theoretic component of the CTCA algorithm.

The random time interval of $t\in [0,T_1]$ is used to introduce
randomness in the order in which sensor nodes perform their power
adjustment routines. Time $T_2$ in line 25 is needed because the
energy level on a node is constantly changing and it helps to insert
this waiting period to modulate how frequently a node requests energy
level information from its reverse-link neighbors and how frequently it broadcasts
its current energy level in response to requests from its reachable
neighbors. Therefore, to ensure that node $N_i$'s reachable neighbors have a
relatively up-to-date information on node $N_i$'s energy level, node
$N_i$ changes its {\em EnergyInfoShared} flag to {\em False} so that
once its reachable neighbors request its information, it will send back the latest
energy level. On the other hand, node $N_i$ should reduce the number
of times that its information is sent due to its own energy
concerns. Therefore, if node $N_i$'s energy level has not changed
noticeably so as to affect its reachable neighbors' actions, it will keep its
{\em EnergyInfoShared} as {\em True} until $T_2$ time has passed. Time $T_3$
in line 27 is needed to ensure that another round of the topology control
process will not begin until the ongoing topology control process has
finished.

\begin{algorithm}[!t]
\caption{The {\em Neighbor-Assisted Power Adjust (NAPA)} function executed
at node $N_i$}\label{TC:game}
\KwIn{current power level $p_i$, current execution number $q$}
\KwOut{new power level $p_i$, execution number $q$}
\If (\tcc*[h]{$Q$ is the maximum number of times a
  node may execute this function per round}) {$q<Q$}
	{$q\leftarrow q+1$\;
	\For {$N_j \in I_i$}
		{\eIf {$S_j$}
			{$k\leftarrow$ index of $p_j$ in $A_j$\;
			$L'_j(\mathbf{P}_i)\leftarrow A_j[k-1]$\;}		
			{$L'_j(\mathbf{P}_i)\leftarrow p_j$\;
			}
		}
	Compute node $m(i)\leftarrow arg \min(L'_j(\mathbf{P}_i)\,|\,N_j\in I_i)$\;
	$S_m\leftarrow$ received $S_m$ from node $m(i)$\;
 	$p_m\leftarrow$ received node $m(i)$'s current transmission power\;
 	$S_i\leftarrow$ {\em AbleToReducePower}$(N_i,p_i)$ \;
 	{\em CanHelp} $\leftarrow$ False\;
	\If {not $S_m$ and $p_m>\min(A_m)$}
 		{$N_{c(m)}\leftarrow N_x\,|\,p(m(i),N_x)=p_m$\;
		\If {$N_{c(m)} \in H_i$ and ${W_i}/{p(N_i,N_{c(m)})}>L'_{m(i)}(\mathbf{P}_i)$}
 			{$p_i\leftarrow p(N_i,N_{c(m)})$\;
  			Broadcast NeighborInfo Request with power $p_i$\;
  			Receive response from newly added neighbors\;
  			Update $R_i$\;
  			$S_i\leftarrow$ {\em AbleToReducePower}$(N_i,p_i)$\;
 			Inform $N_j\in \{R_i\cup I_i\}$ of $(p_i,S_i)$\;
  			{\em CanHelp} $\leftarrow$ True\;
			}
		}
	\If {not {\em CanHelp}}
		{\If {$p_i>p'_i(\mathbf{P}_i)$}
			{$p_{\text{temp}}\leftarrow p_i$\;
  			$p_i\leftarrow p'_i(\mathbf{P}_i)$\;
  			Update $R_i$.\;
 			$S_i\leftarrow$ {\em AbleToReducePower}$(N_i,p_i)$\;
 			Broadcast $(p_i,S_i)$ with power $p_{\text{temp}}$\;
 			Inform $N_j\in I_i$ of current $(p_i,S_i)$\;
			}
		}
	\eIf {$S_i$}
		{ Wait for a random time $t\in [0,T_1]$\;
 		$p_i,q\leftarrow$ NAPA($p_i,q$)\;}
 		{{\bf return} $p_i,q$\;
		}
}
\end{algorithm}

\begin{algorithm}[t]
\caption{The {\em AbleToReducePower} function executed at node $N_i$}\label{TC:potentail_lifetime}
\KwIn{current power level $p_i$}
\KwOut{Able to reduce power flag $S_i$}
 $N_{c(i)}\leftarrow N_x\,|\,p(N_i,N_x)=p_i$\;
 $W_{c(i)} \leftarrow$ node $N_{c(i)}$'s current power level\;
  $S_i\leftarrow$False\;
 $p'_i(\mathbf{P}_i)\leftarrow p_i$\;
\If {$p_i>\min(A_i)$}
	{\For {$N_j\in R_i$}
		{\If {$N_{c(i)}\in R_j$ and $L_i(p_i)<\frac{W_{c(i)}}{p(N_{c(i)},N_j)}$}
			 {$S_i\leftarrow$ True\;
			 $k\leftarrow$ index of $p_i$ in $A_i$\;
			 $p'_i(\mathbf{P}_i)\leftarrow A_i[k-1]$\;
 			\textbf{break}\;
			}
		}
}
 {\bf return} $S_i$\;
\end{algorithm}

The detailed {\em NAPA} function is illustrated in
Algorithm \ref{TC:game}. As we have explained in Section
\ref{sec:analysis}, each node should always try to meet its primary
goal unless it cannot be
accomplished. This process for helping improving its primary goal is
illustrated in lines 16--26 in Algorithm \ref{TC:game}. If
$N_i$ is to increase its transmission power to help improve its
reverse-link neighborhood's potential lifetime (as illustrated by node $N_3$ in
Fig.\,\,\ref{fig:2-after-1}), several conditions have to be met:
\begin{enumerate}
\item The node with the minimum potential lifetime within $N_i$'s reverse-link
neighborhood (node $m(i)$) cannot improve its potential
lifetime on its own ($S_m$ is False), as in node $N_1$'s case illustrated in
Fig.\,\,\ref{fig:2-before}.
\item Node $m(i)$ is not transmitting at its minimum transmission power ($p_m>\min(A_m)$).
\item $m(i)$'s potential lifetime can be improved with $N_i$
  transmitting at a certain larger power $a_i$. In this case,
  $N_{c(m)}\in H_i$, indicating that node $N_i$ should be transmitting
  at power $p(N_i,N_{c(m)})$.
\item $N_i$'s lifetime when transmitting at power $p(N_i,N_{c(m)})$ is
  larger than its reverse-link neighborhood's potential lifetime,
  i.e., $W_i/p(N_i,N_{c(m)})>L'_{m(i)}(\mathbf{P}_i)$.
\end{enumerate}

Conditions (1) and (2) are implemented in line 16 of
Algorithm \ref{TC:game}, and conditions (3) and (4) are implemented in line
18 in Algorithm \ref{TC:game}. If all of the conditions listed above have
been met, then node $N_i$ will choose to increase its transmission
power so as to help improve its reverse-link neighborhood's potential
lifetime. On the other hand, if the node cannot
help to improve its reverse-link neighborhood's potential lifetime (indicated by {\em
CanHelp} being False in line 28), then it will try to meet its
secondary goal of improving its own estimated lifetime. This process
is indicated by lines 29--36. In cases where a node can still
improve its estimated lifetime, it will schedule to perform the NAPA
function again after a random period of time (lines 39--40).

Function {\em AbleToReducePower} in Algorithm \ref{TC:potentail_lifetime} illustrates the procedure
implemented by each node to calculate its potential transmission
power.
It is also the function that helps a node $N_i$ determine its local connectivity function $c_i(a_i,t)$ at power level $a_i$.
If there exists a reachable neighboring node $N_j$ of $N_i$ such that
$N_j$ can communicate with the node that determines $N_i$'s current
transmission power 
(denoted by $N_x$),
then $N_i$'s potential transmission power, denoted by $p'_i(\mathbf{P}_i)$, is one
level below its current transmission power, and $S_i$ is True. 
In other words, there exists a path between node $N_i$ and node $N_x$
when node $N_i$ is transmitting at power level $p'_i(\mathbf{P}_i)$,
and thus, $c_i(p'_i(\mathbf{P}_i),t)=1$. For any power level
$a_i>p'_i(\mathbf{P}_i)$, we have $c_i(a_i),t)=1$. On the other hand,
if such node $N_j$ could not be found, we have
$p'_i(\mathbf{P}_i)=p_i$, and $S_i$ is False. At this point, for any
power level $a_i<p_i$ and $c_i(a_i,t)=0$. This is because at power level
$a_i<p_i$, node $N_i$ is transmitting at a power level lower than what
is necessary to stay connected with node $N_x$, and therefore, loses
its local connectivity.  Note that, in either of the two cases, no
communication among sensor nodes has to be conducted in order to
calculate node $N_i$'s local connectivity function $c_i(a_i,t)$. 

\begin{algorithm}[!t]
\caption{Functions executed at $N_i$ upon receiving
control/request messages.} \label{TC:upon_request}
\textbf{Upon receiving transmission power updates from $N_j$}\\
\If {$N_j\in I_i$}
	{\eIf {$p_j<p(N_j,N_i)$}
		{Remove $N_j$ from $I_i$\;}{Update $N_j$'s entry in $I_i$ with $(p_j,S_j)$\;}
}
\If {$N_j \in R_i$}
	{Update $N_j$'s entry in $R_i$ with $(p_j,S_j)$\;
	$S_{\text{temp}}\leftarrow S_i$\;
	$S_i\leftarrow$ {\em AbleToReducePower}$(N_i,p_i)$\;
	\If {$S_i\neq S_{\text{temp}}$}
		{Broadcast $S_i$ with power $p_i$\;
		Wait for a random time $t\in [0,T_1]$\;
		$p_i,q \leftarrow$ {\em NAPA}($p_i,q$)\;}
}
 \BlankLine
\textbf{Upon receiving {\em NeighborInfo} Request from $N_j$}\\
\If {$N_j\not\in I_i$}
	{Inform $N_j$ of $(W_i,p_i,S_i)$\;
	Add $N_j$ to $I_i$\;}
 \BlankLine
\textbf{Upon receiving remaining energy request}\\
\If {not {\em EnergyInfoShared}}
{ Broadcast $W_i$ with power level $p_i$\;
{\em EnergyInfoShared} $\leftarrow$ True\;
}
\end{algorithm}

To ensure up-to-date information sharing amongst a node's reachable
neighborhood and its reverse-link neighborhood, the communication
routines that are executed by each node are illustrated in
Algorithm \ref{TC:upon_request}. These routines ensure that once a
node has changed its current status (such as current transmission
power, potential transmission power and current remaining energy),
nodes whose status may be affected are informed.

As has been proved in the previous section, game $\Gamma=\langle
N,A,U\rangle$ is an ordinal potential game, and seeking the optimal
global potential function yields a Nash equilibrium. Therefore, given
enough time, the NAPA procedure converges to an equilibrium. In our
observations, we find that $Q=4$ is adequate to ensure good
performance. Therefore, in our implementations, we allow only four
executions of the {\em NAPA} function per round per node.

The initialization stage of the CTCA algorithm as illustrated in
Fig.\,\,\ref{TC:CTCA} introduces the same order of computational and
communication complexity as the DLSS algorithm, which is
$O(\Delta^2)$. The communication and computation complexity of the
CTCA algorithm at each round is $O(\Delta)$.

\section{Simulation Results}\label{sec:CTCA_performance}

\subsection{Simulation and Energy Consumption Model}

The energy model used in our simulation is identical to that used in
the research literature on topology control
\cite{HeiCha2002,KolPav2011}. This model incorporates energy
consumption due to transmission, reception, and for radio electronics
in both free space and over a multi-path channel above a certain
distance threshold.
\begin{eqnarray}
E_{T_x}(d) & = &E_{elec}\times k+\left\{
   \begin{aligned}
   \varepsilon_{fs} \times d^2\times k \text{\ \ if } d<d_0\\
   \varepsilon_{mp} \times d^4\times k \text{\ \ if }  d\geq d_0 \\
   \end{aligned}
   \right.\nonumber\\
E_{R_x} & =  & E_{elec}\times k \label{equ:original}
\end{eqnarray}
where $E_{T_x}(d)$ is the energy consumed in transmitting the signal
to an area of radius $d$ and $E_{elec}$ is the energy consumed for
the radio electronics. $\varepsilon_{fs}$ is the transmitter's
amplifier coefficient in free space and $\varepsilon_{mp} $ is the
transmitter's amplifier coefficient in the multi-path channel. $d_0$
is the distance threshold beyond which the channel is considered as
multi-path. $E_{R_x}$ is the energy consumed in receiving the
signal, and $k$ is the number of bits in the packet. Radio
parameters are set as $E_{elec}=50 nJ/bit$,
$\varepsilon_{fs}=10pJ/bit/m^2$,
$\varepsilon_{mp}=0.0013pJ/bit/m^4$, and $d_0=87.8m$.

Our simulation is conducted for a square 10km$\times$10km region
within which 200 nodes are placed in random locations. Each node is
equipped with 40kJ of energy and has a maximum transmission power
$p_{\text{max}}$, which corresponds to a transmission radius of 20\%
of the width of the square region. The constant $T_3$ in the CTCA
algorithm is chosen to be $1000$ times larger than $T_1$, and $T_2$ is
chosen to be half of $T_3$. Each data point in the results reported
here is the average of 200 randomly generated graphs. Using the batch
means method to estimate confidence intervals, we have determined that
the 95\% confidence interval is within $\pm$2\% for each of the data
points reported in our results. In our simulation model, we employ the
TinyOS standard \cite{tinyos} for sensor node data transmission,
including its packet formats.

\begin{figure*}[!t]
\begin{center}
   {
       \includegraphics[width=3.5in]{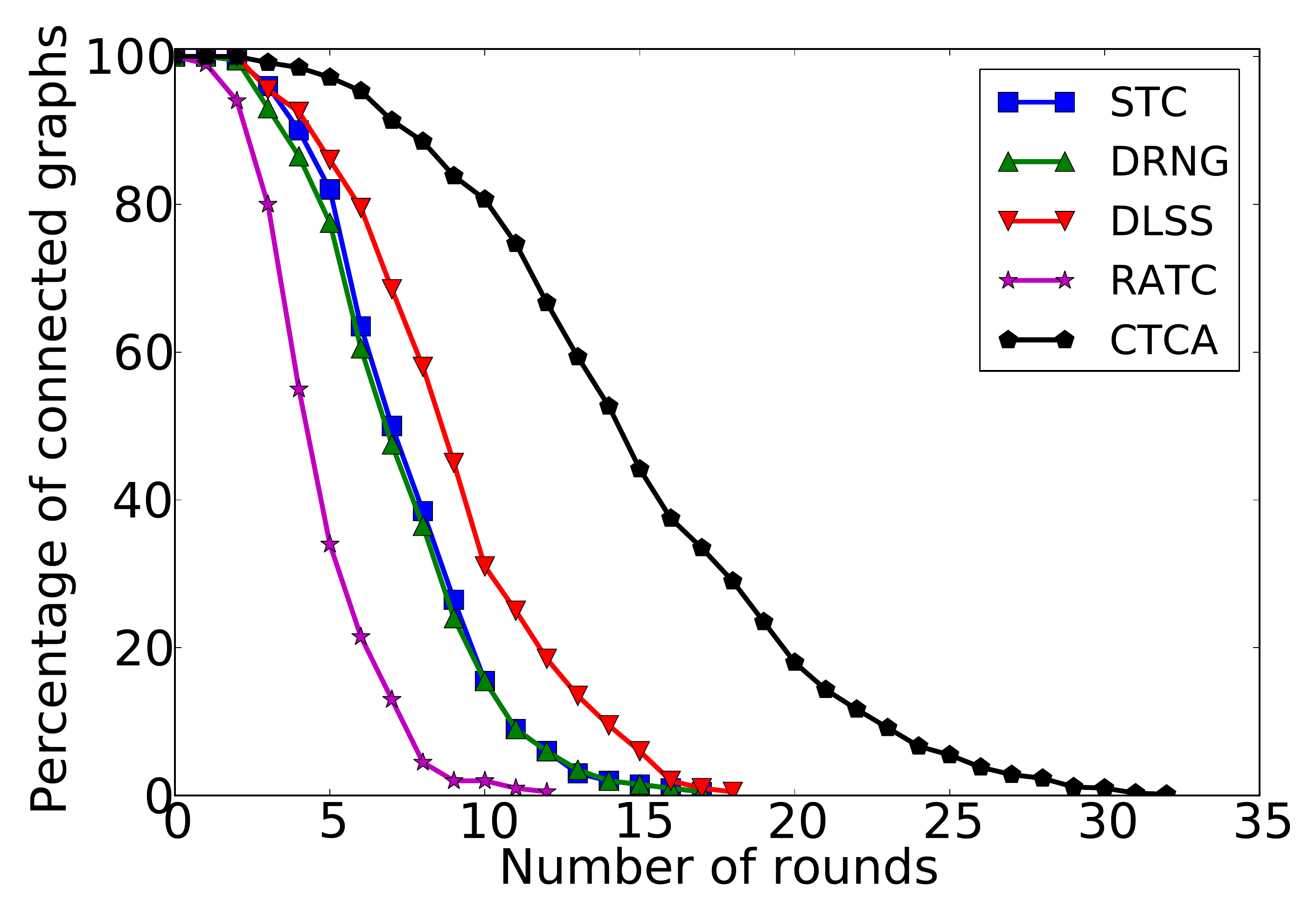}
       }
   \caption{Lifetime comparison against different algorithms}\label{fig:lifetime}
\end{center}
\end{figure*}

\begin{figure*}[!t]
\begin{center}
    \subfigure[{The average transmission power per node plotted against time until 50\% of the graphs in the
      simulation experiments lose connectivity.}]{
        \includegraphics[width=3.5in]{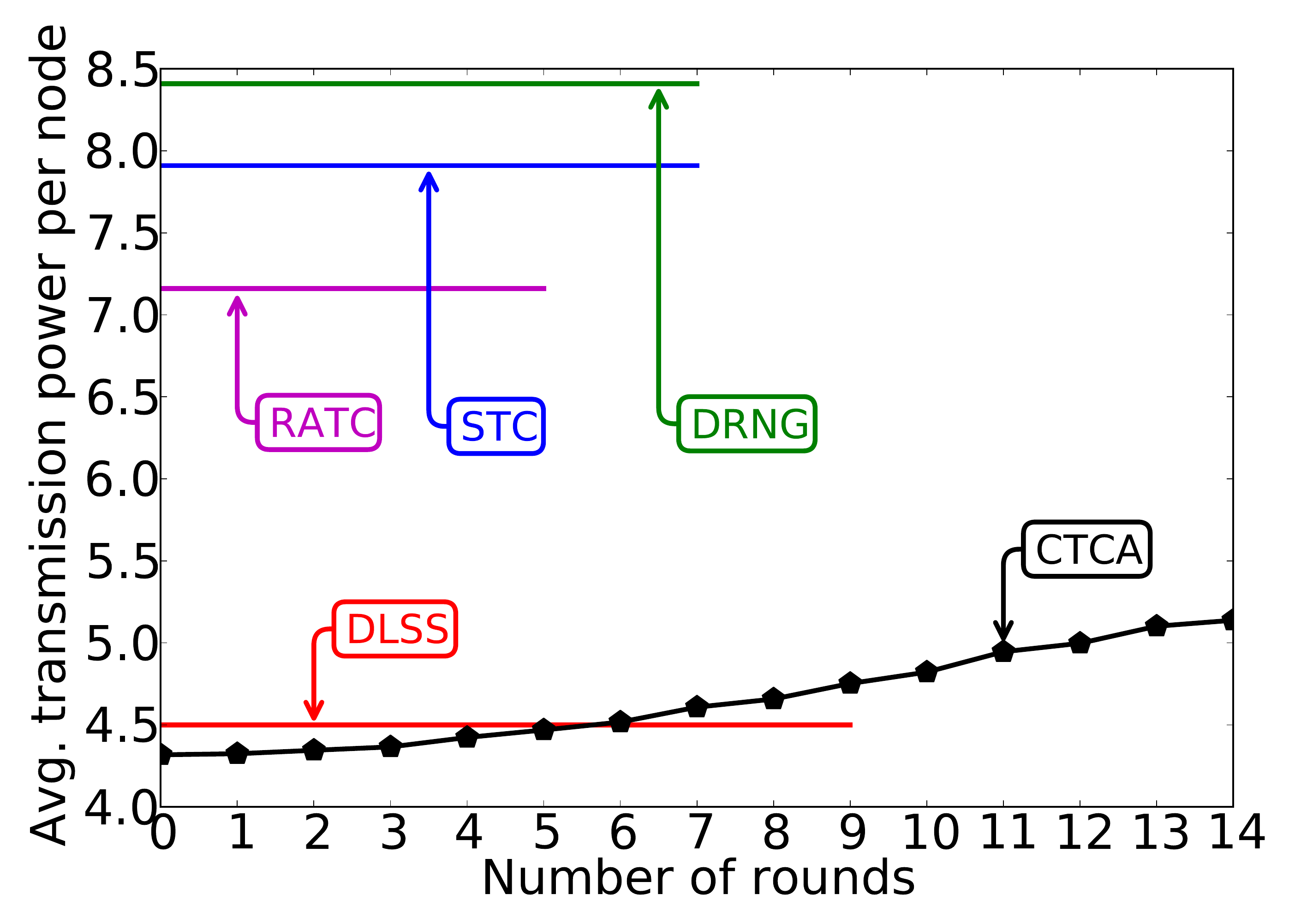}
        \label{fig:power}
        }\\
    \subfigure[{The average energy consumption along the minimum
      energy path plotted against time until 50\% of the graphs in the
      experiments lose connectivity.}]{
      \hskip-0.55cm\includegraphics[width=3.75in]{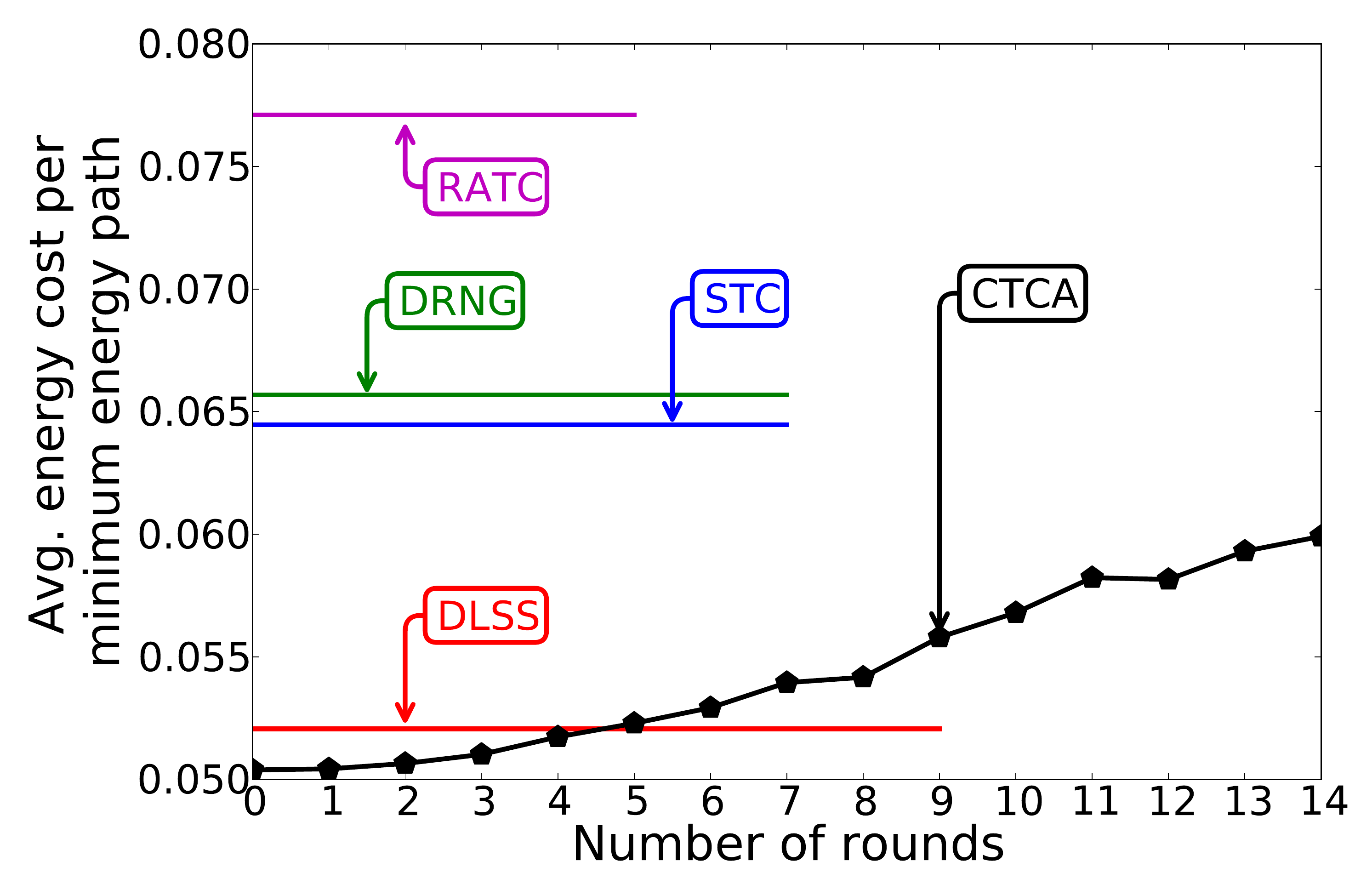}
        \label{fig:cost}
        }
   \caption{A comparative performance analysis of different algorithms. }\label{fig:Type1}
\end{center}
\end{figure*}

\subsection{Comparative analysis against other algorithms}
In this section, we compare the performance of CTCA algorithm
against some of the other algorithms. Among the well-cited algorithms, our
criteria for including them in this comparative analysis are the following:
\begin{itemize}
\item The algorithm applies to or allows application in which communication is uni-directional, where if node $N_i$ is within the communication radius of node $N_j$, node $N_j$ is not required to be within the communication radius of node $N_i$. It is the same assumption that we make in this paper.
\item The communication and computational complexity for an adaptive
algorithm is $O(\Delta)$ or lower each round.
\end{itemize}

Based on the above criteria, we have selected {\em Directed Relative
  Neighborhood Graph (DRNG)} \cite{LiHou2005-1313}, {\em Directed Local
    Spanning Subgraph (DLSS)} \cite{LiHou2005-1313}, {\em Step Topology
      Control (STC)} \cite{SetGer2010}, and {\em Routing Assisted Topology
      Control (RATC)} \cite{KomMac2009}. In the case of the RATC
    algorithm, it was reported in \cite{KomMac2009} that when sensor nodes operate under a
given level of 3-hop knowledge, the algorithm yields the best performance. Thus, we also allow up to 3-hop level of information to be exchanged among sensor nodes in the RATC algorithm.

In our experiments, every round, each node will send a designated
data packet to every other node within the network, i.e, a node will
send out $n-1$ packets each round. Data packets are routed through the
minimum energy consumption path. In case of the CTCA algorithm, at
the beginning of each round, each node adjusts its transmission
power according to the energy situation in its local area; as for all other
algorithms, each node will send out a hello message to check their
neighbors' availability.

In our simulations, we include the full energy costs of the overhead
(such as hello messages) associated with executing each of the
algorithms considered in the comparative analysis.


Fig.\,\,\ref{fig:lifetime} reports the network lifetime achieved by
the different algorithms. For each point in the graph, its x-axis value
indicates the number of rounds that has passed. The y-axis
value indicates the percentage of graphs (of the 200 randomly
generated graphs used as a starting point in the experiments) that are
still connected. As shown in the figure, a significantly larger
fraction of graphs stay connected when using the CTCA compared to
other previously-known algorithms. On average, we find that the life
of a network is extended by more than 50\% compared to other
algorithms.

Fig.\,\,\ref{fig:Type1} reports the network's parameters (the average
transmission power per node and the average energy cost along the
minimum energy path) achieved by the different algorithms. 
For static algorithms such as DLSS,
DRNG, STC, and RATC, the topology of the network is determined at the
very beginning of the network's lifetime. Therefore, their network's
parameters remain the same throughout the network's lifetime, as
indicated by straight lines in Figs. \ref{fig:power} and \ref{fig:cost}. The CTCA
algorithm, on the other hand, changes the topology of the network
with time, and thus, produces different parameters each round. In
Figs. \ref{fig:power} and \ref{fig:cost}, we have reported each algorithm's performance
until 50\% of the random graphs that we have generated become
disconnected. DLSS is the only algorithm that achieves average
transmission power or energy cost per path comparable to the CTCA
algorithm. However, as time progresses, all algorithms except CTCA
retain the same average transmission power per node until the network's
functional life ends, but the CTCA algorithm adapts accordingly and
preserves connectivity for much longer. It is worth noting that, in
the case of the CTCA algorithm, between the first round when a graph
is connected to the 14th round when it is only 50\% likely that it is
connected, the average transmission power per node in the CTCA
algorithm increases by only about 20\%. The same observation can be
made for the average energy cost along the minimum energy path.

Note that, the CTCA algorithm is an algorithm that determines the
topology of the network. Therefore, in our simulation, to capture how
the CTCA algorithm is able to help extend the network's functional
lifetime, we only employ the simplest routing algorithm---routing the
packet through the minimum energy path. The performance of the CTCA
algorithm may vary depending upon the routing algorithm used to
transmit data packets. An efficient routing algorithm may help extend
the network's functional lifetime even further if the energy
dissipation can be distributed more evenly. 
However, whether or not
the routing algorithm is an efficient one, the CTCA algorithm
accommodates the impact of the routing algorithm because it 
dynamically adapts to the current energy level at each node. 


\subsection{Comparison against the optimal solution}

In topology control algorithms, the weight of an edge usually
reflects the cost of transmission through that particular
link. Alternatively, if we assign the weight of an edge $(N_i,N_j)$ at
time $t$ as:
\begin{equation} w_{i,j}(t)=\frac{p(N_i,N_j)}{W_i(t)}\label{eqn:weightfunction}
\end{equation}
then, this weight function captures the amount of estimated lifetime
consumed by the sender node $N_i$ if a transmission is made through
link $N_i\rightarrow N_j$. A centralized topology control algorithm to
minimize the maximum weight of an edge while preserving connectivity
is trivial (e.g., based on removing edges from the graph in order of
decreasing weight until removing an edge would destroy connectivity). Let
$T_{\text{Opt},r}$ denote the maximum possible estimated lifetime of
the network achieved using this optimal algorithm on the input graph
in round $r$.

Let $T_{\text{CTCA},r}$ denote the estimated lifetime of the network
achieved using the CTCA algorithm on the input graph in round $r$. We
define the {\em average price paid by the CTCA algorithm} as:
\[
\text{Average price paid} = \text{avg} \left( \frac{T_{\text{Opt},r}}{T_{\text{CTCA},r}}\right)
\]
In our performance analysis, we compute the above average price paid
by taking the average over multiple runs each with a different random
graph as the input. We use the term {\em
  price} in line with the traditional terminology in game theory used
in metrics comparing Nash equilibrium solutions against the social
optimum (e.g., the price of anarchy \cite{KouPap1999}). In this
section, we will use this metric (the average price paid by CTCA) as a
measure of the quality of the solution reached by the CTCA algorithm
(note that this only measures the quality of the solution and not the
energy expended to reach the solution, addressed in the previous
subsection in computing actual lifetimes).

\begin{figure}[!t]
\begin{center}
    \subfigure[{The average price paid by CTCA (average ratio of
      the optimal lifetime and that achieved by CTCA).}]{
       \includegraphics[width=3.5in]{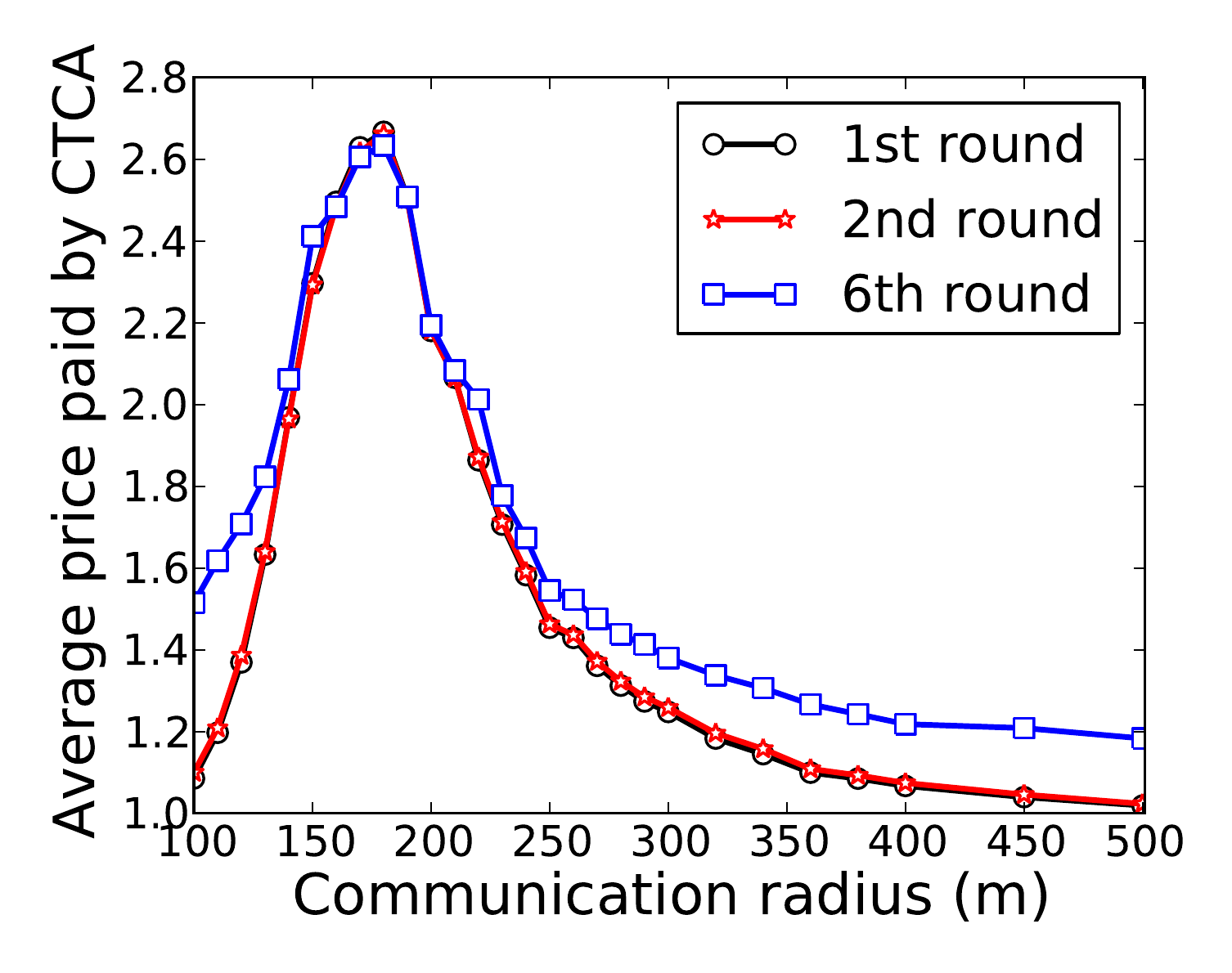}
       \label{fig:perf_range_price}
       }\\
    \subfigure[{The percentage of times that the CTCA algorithm finds
      the optimal solution.}]{
       \hskip-0.5cm \includegraphics[width=3.7in]{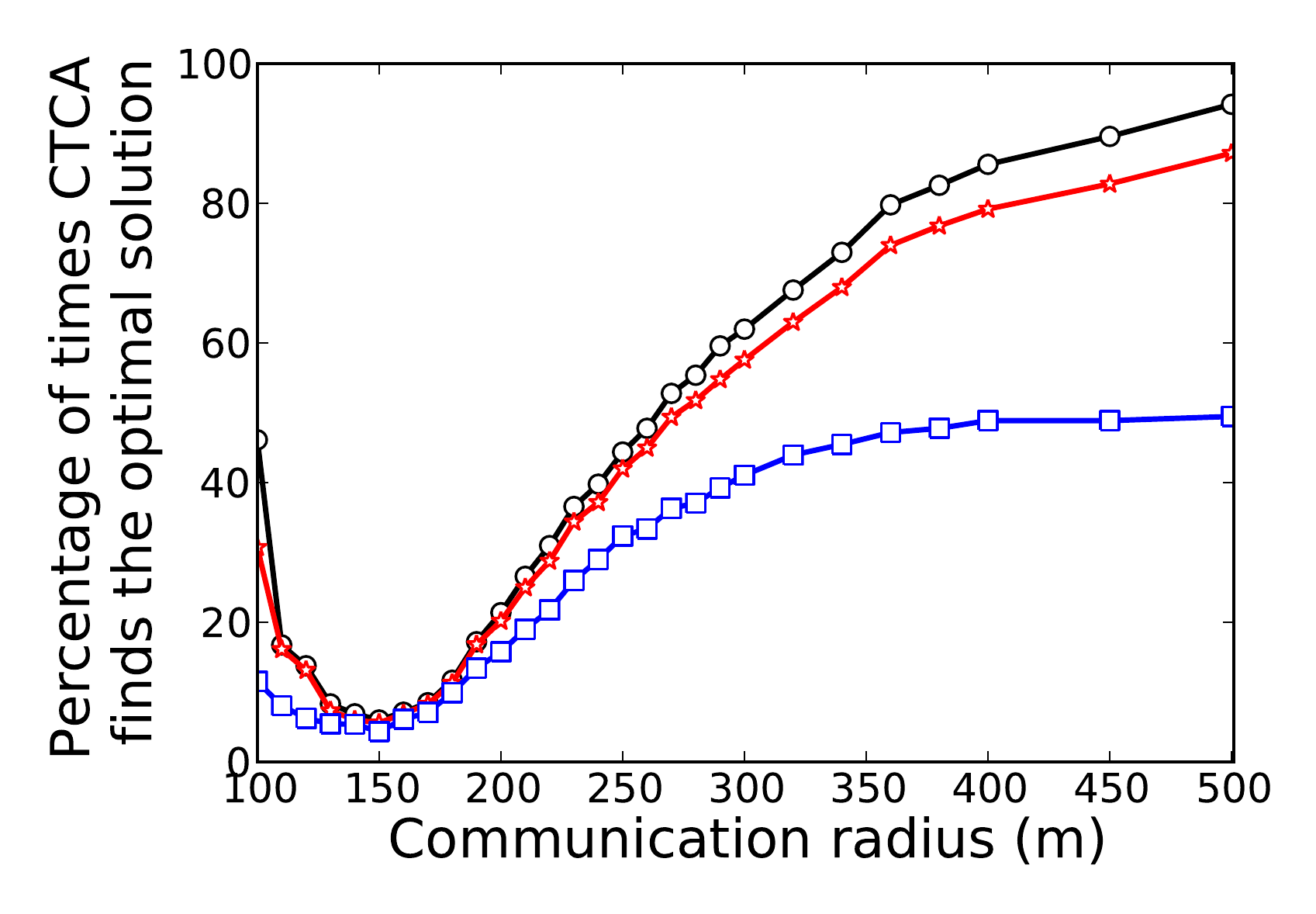}
        \label{fig:perf_range_percent}
        }
   \caption{The performance of the {\em distributed} CTCA algorithm in
     comparison to the {\em centralized} optimal algorithm in
     different rounds plotted against the communication radius of the nodes.}\label{fig:perf_range}
\end{center}
\vspace{-12pt}
\end{figure}

In the following set of experiments, we study the quality of the
solution delivered by the distributed CTCA algorithm in comparison to
the centralized optimal solution. In our experiments, we begin with
sensor nodes all with the same amount of starting energy (10J) and
randomly deployed in a 1000m $\times$ 1000m square region. While the
energy levels of nodes at the beginning of the 1st round is the same,
the uneven distribution of the energy consumption on the sensor nodes
leads to unevenness in the energy levels of the nodes in subsequent
rounds. Accommodating this unevenness being an important goal of this
paper, we trace the performance of the CTCA algorithm in different
rounds (1st, 2nd and 6th). We choose the first round because it is
when energy levels are all the same. We choose the 2nd round because
this is the first round at the beginning of which energy levels may be
different on different nodes. We choose the 6th round because, as
shown in Fig. \ref{fig:lifetime}, the network may have passed nearly
50\% of its lifetime after this many rounds.

\subsubsection{The influence of communication radius of nodes}\label{sec:radius}

In our first set of experiments, we study the influence of the
sensor node's maximum communication radius on the performance of the
CTCA algorithm. We use 200 sensor nodes deployed in the region. We
conduct 500 independent simulations and report the results in
Fig. \ref{fig:perf_range}. Fig. \ref{fig:perf_range_price} reports
the average price paid by the CTCA algorithm with different values of
the communication radius of nodes in each of rounds 1, 2 and
6. Fig. \ref{fig:perf_range_percent} reports the percentage of times
that the CTCA algorithm is able to find the optimal solution for
different communication radii in those same rounds.

Fig. \ref{fig:perf_range_price} shows that when the communication
radius is very small (e.g., at 100 meters or 10\% of the length of each
side in the square area), the performance of the CTCA algorithm is
very close to that of the optimal algorithm (with average price paid
close to 1.0). This is because, given the same sensor node
density, the topology graph is already very sparse when the
communication range of the nodes is relatively small. The CTCA
algorithm as well as the optimal algorithm cannot do much to improve
the lifetime in this situation. On the other hand, as the communication
range of the sensor nodes is increased, the number of choices
available to each sensor node when trying to adjust its transmission
power increases. A centralized algorithm is better able to exploit
these choices because of more information available to it as compared
to the localized information available to each node in the distributed
CTCA algorithm. As the communication range
increases even more, each node is able to gain sufficient information about
the region around itself even in a distributed algorithm like
CTCA. Therefore, at larger communication ranges, the disparity in the
performance of the distributed CTCA algorithm and the centralized
optimal algorithm reduces again with the average price paid by the
CTCA algorithm reaching closer to 1.0.

The above phenomenon also explains why, as shown in
Fig. \ref{fig:perf_range_percent}, the CTCA algorithm is more likely
to find the optimal solution when the communication range of nodes is
very low compared to when the range is intermediate. The figure also
shows that, as expected, the CTCA algorithm finds the optimal solution
with high likelihood when the communication range of the nodes is
high. Note that the weaker performance at intermediate ranges is an
inherent limitation of a distributed algorithm which works with only
localized information and not necessarily of the CTCA algorithm
(which performs better than other distributed algorithms as shown in
the previous subsection).

\begin{figure}[!t]
\begin{center}
    \subfigure[{The average price paid by CTCA (average ratio of
      the optimal lifetime and that achieved by CTCA).}]{
       \includegraphics[width=3.5in]{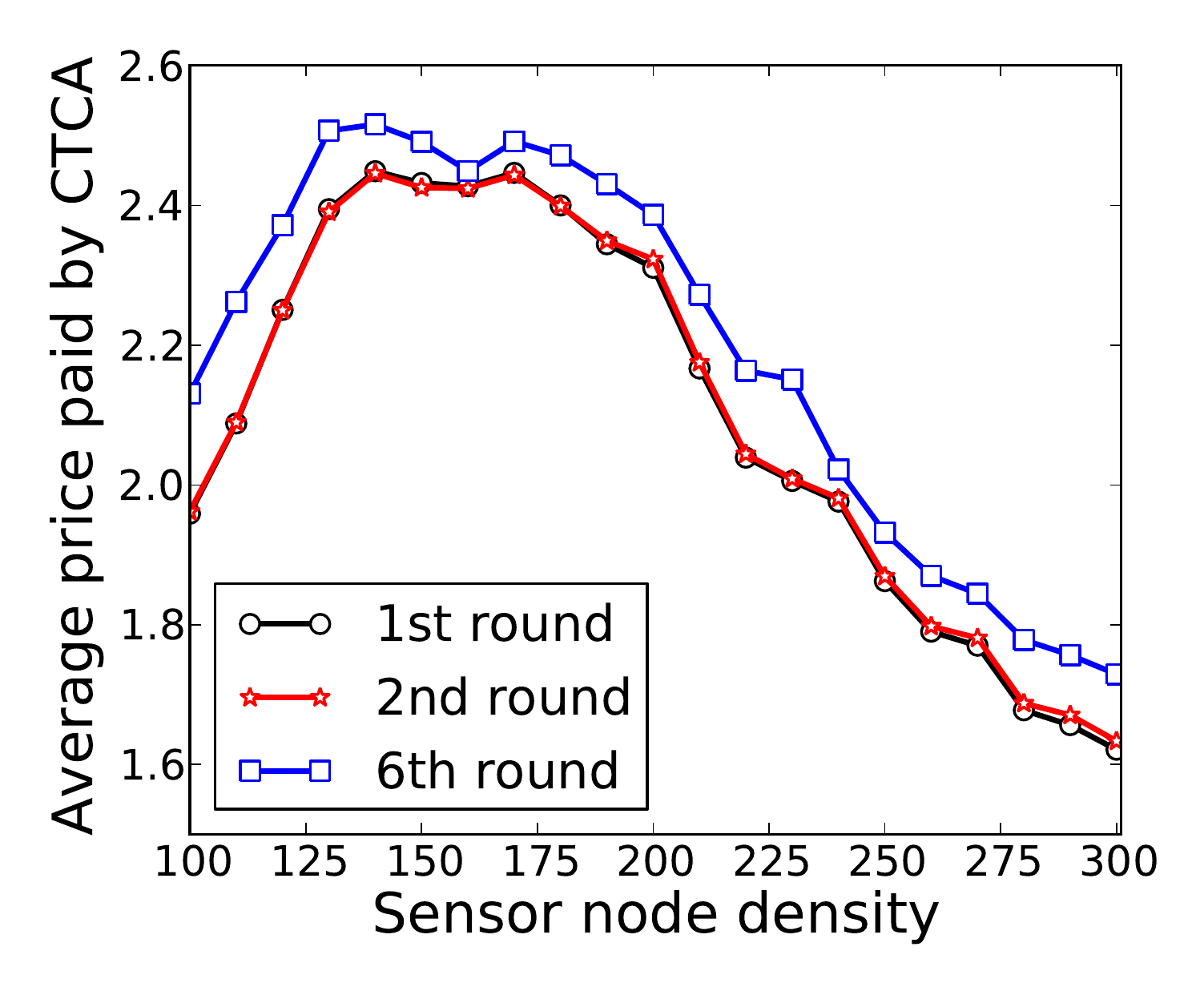}
       \label{fig:perf_density_price}
       }\\
    \subfigure[{The percentage of times that the CTCA algorithm finds
      the optimal solution.}]{
        \hskip-0.5cm\includegraphics[width=3.65in]{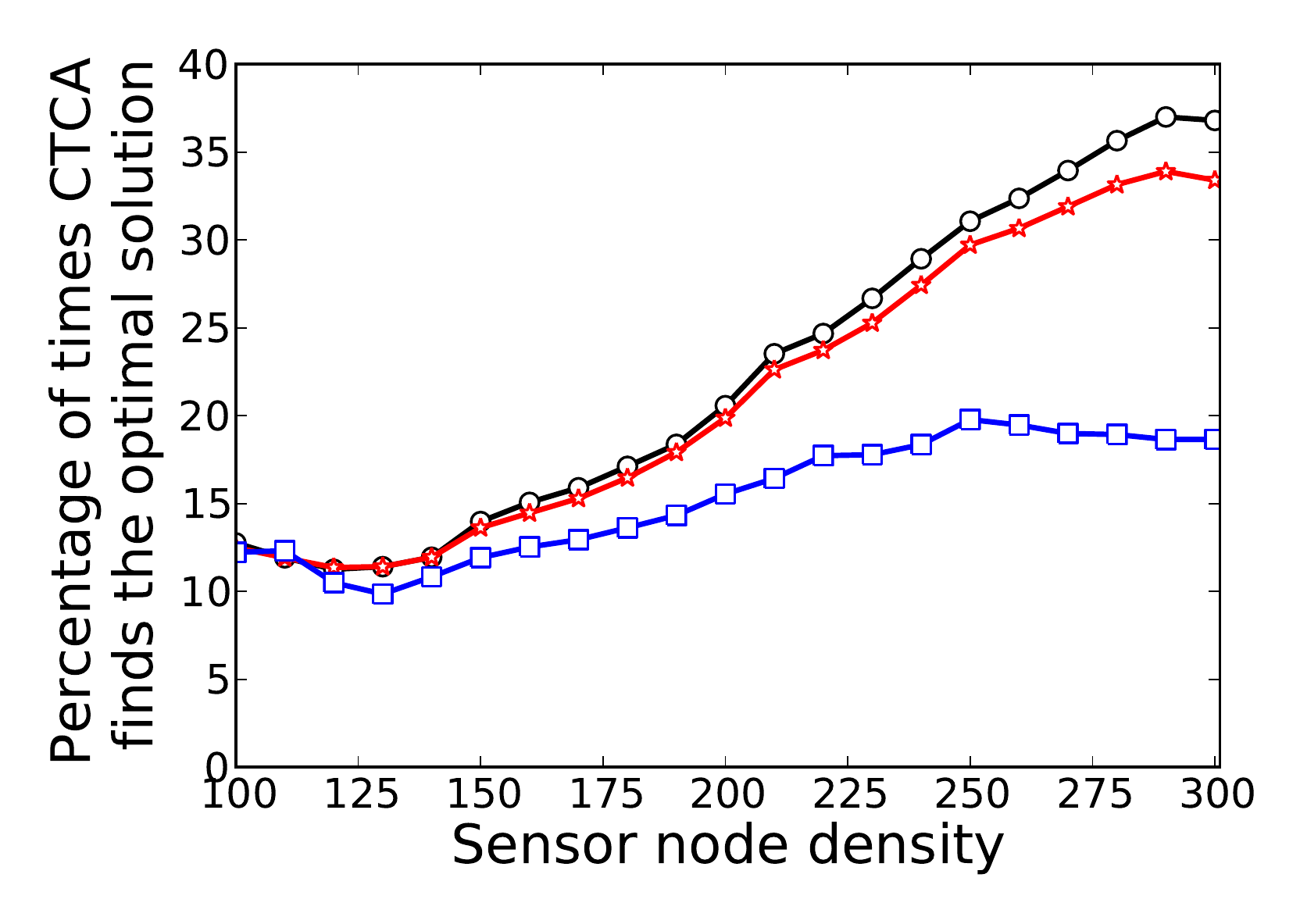}
        \label{fig:perf_density_percent}
        }
   \caption{The performance of the {\em distributed} CTCA algorithm in
     comparison to the {\em centralized} optimal algorithm in
     different rounds plotted against sensor node density, expressed
     in number of nodes per square kilometer.}\label{fig:perf_density}
\end{center}
\vspace{-12pt}
\end{figure}

It is of interest to observe that, in round 6, the disparity in the
energy levels remaining on different nodes becomes larger and which, in
turn, triggers more frequent cooperation between nodes and more
changes in the topology. Depending on the order in which different
nodes cause these changes, the potential solution space available to
the CTCA algorithm can be large. As a result, even though the average
performance of the CTCA algorithm in the 6th round is similar to that
in other rounds, it is much less likely to reach the perfect optimal
solution in the 6th round.

\subsubsection{The influence of sensor node density}\label{sec:density}

In our second set of experiments, we limit the maximum communication
radius of each sensor node to 200 meters (when the CTCA algorithm
performs close to the worst in comparison to the optimal centralized
algorithm). To study the impact of the sensor node density on the
performance of the CTCA algorithm, we conducted an experiment with
sensor node densities ranging from 100 to 300 per square kilometer
(i.e., 100 to 300 nodes in the region in our simulation
experiments). The results are reported in
Fig. \ref{fig:perf_density}. Fig. \ref{fig:perf_density_price}
reports the average price paid by the CTCA algorithm for different
densities in different rounds. Fig. \ref{fig:perf_density_percent}
reports the percentage of times that the CTCA algorithm is able to
find the optimal solution.

In Fig. \ref{fig:perf_density}, we observe a similar trend as in
Fig. \ref{fig:perf_range} with a dip in the performance at
intermediate communication ranges in the former figure and a dip in the performance
at intermediate densities in the latter figure. The trend is explained
by the same phenomena described earlier in the context of changes in
performance with changes in the communication radius of the nodes.


\section{Conclusion}\label{sec:conclusion}
In this paper, we proposed a game-theoretic approach for nodes in a
sensor network to cooperatively change their transmission powers to
help extend the network lifetime. We have proved the existence
of a Nash equilibrium for our game and provided an algorithm,
Cooperative Topology Control with Adaptation (CTCA), which achieves
such an equilibrium. Our simulation results show that the CTCA
algorithm is able to improve the lifetime of a wireless sensor network
by more than 50\% compared to the best previously-known algorithms.

To better assess the performance of the CTCA algorithm, we also
compare the quality of the topology delivered by the CTCA algorithm to
the optimal solution obtained using a centralized algorithm. Our
results show that with increased information available to each node
about its region (such as when the communication range is large), the
CTCA algorithm performs closer to the optimal one. Also, the more
topological options available to each node (such as when the node
density is high), the more likely that the average performance of the
CTCA algorithm is closer to the optimal one.

While a distributed algorithm like the CTCA is able to perform well
with more information or options available at each node, we find that
there is a significant gap between the average performance of the CTCA
algorithm and the optimal centralized one. Even though the CTCA
algorithm performs better than other distributed algorithms, this
paper suggests that there may yet be more room for new research on
better distributed algorithms.

While our work has used a game-theoretic approach under the
constraint that the network remain connected, our algorithm can be
adapted to other criteria that describe the functional life of a
network (such as whether or not each portion of a certain region
is covered by a sensor node within a pre-defined distance). The
connectivity is captured in the term $C(\mathbf{P})$ in Equation
(\ref{potential}) and can be replaced by a different criterion such as
coverage. Our ongoing research is focused on developing and describing
a generalized version of this approach.
\bibliographystyle{IEEETran}
\bibliography{chu-sethu}
\end{document}